\documentclass[prd, 10pt,aps,nofootinbib, floatfix, notitlepage,twocolumn]{revtex4-1}

\usepackage{amsmath,amsfonts,amssymb}
\usepackage{mathrsfs}
\usepackage{graphicx}
\usepackage[english]{babel} 
\usepackage{url} 
\usepackage{color}
\usepackage{bbold}
\usepackage{adjustbox}
\usepackage[dvipsnames]{xcolor}
 \usepackage[utf8]{inputenc} 
\usepackage{float}
\usepackage[colorlinks=true,citecolor=blue,urlcolor=blue]{hyperref}
\usepackage[noabbrev]{cleveref}
\usepackage{csquotes}
\usepackage[normalem]{ulem}
\usepackage{aas_macros}
\usepackage{xspace}
\usepackage{subfigure}

\newcommand{\be}{\begin{equation}}
\newcommand{\ee}{\end{equation}}
\newcommand{\bes}{\begin{equation*}}
\newcommand{\ees}{\end{equation*}}

\newcommand{\LCDM}{$\Lambda$CDM\xspace}

\makeatletter
\def\l@subsubsection#1#2{}
\makeatother

\begin{document}

\title{Predictions for new physics in the CMB damping tail}

\author{Tristan L. Smith${}^{a}$}
\email{tsmith2@swarthmore.edu}
\author{Nils Sch\"oneberg${}^{b,c}$}
\email{nils.science@gmail.com}
\affiliation{${}^{a}$Department of Physics and Astronomy, Swarthmore College, Swarthmore, PA 19081, USA}
\affiliation{${}^{b}$University Observatory, Ludwig-Maximilians-Universität, Scheinerstr.
1, 81677 Munich, Germany}
\affiliation{${}^{c}$Excellence Cluster ORIGINS, Boltzmannstr. 2, 85748 Garching, Germany}
\date{\today}

\begin{abstract}
Ever since the \emph{Planck} satellite measured the the cosmic microwave background (CMB) down to arcminute angular scales, the mismatch between the CMB-inferred value of the Hubble constant and the value inferred from the distance ladder (i.e., the Hubble tension) has been a growing concern and is currently at the $\sim 6 \sigma$ level. There are a handful of proposed mechanisms operating in the early universe which have shown some promise in resolving the Hubble tension. These mechanisms are expected to leave a measurable impact on the smallest scale CMB anisotropy, deep in the damping tail. Using current CMB data, baryonic acoustic oscillation data, and the luminosities of Type Ia supernovae as a baseline, we compute the predicted small-scale CMB power spectra for a characteristic set of these models. We find that near-future CMB data should be able to distinguish some but not all of the investigated models from the core cosmological model, $\Lambda$CDM.
\end{abstract}

\maketitle

\section{Introduction}

For around three decades the $\Lambda$CDM model has been firmly established as the baseline cosmological model. It consists of baryons, photons, neutrinos, cold dark matter, and a cosmological constant ($\Lambda$) required to explain the late time accelerated expansion of the universe. During this time it has been repeatedly subjected to a variety of cosmological data and throughout emerged as a consistent description. Various different cosmological probes, such as the cosmic microwave background (CMB) anisotropies, the clustering of large scale structure (LSS) and in particular the baryon acoustic oscillations (BAO), the Type Ia supernovae (SNeIa) brightness, cosmic chronometers (CC), as well as the primordial abundance of light elements generated during big bang nucleosynthesis (BBN), have progressively increased the precision of the determination of cosmological parameters to the percent level, see for example Ref.~\cite{2020RvMP...92c0501P}. However, while these ever tighter measurements continue to be well-fit in the $\Lambda$CDM model, different data sets prefer discrepant values of some of the cosmological parameters. The most significant of these tensions is the difference between the Hubble constant, $H_0$\,, as measured through late-time direct kinematical measurements and as indirectly inferred through early-time observations. While the value of $H_0$ measured through the calibrated SNeIa distance ladder has consistently remained higher than $70$ km/s/Mpc (see e.g.~Ref.~\cite{Brout:2022vxf,Riess:2021jrx,Freedman:2021ahq}), recent observations of ever-smaller scales in the CMB anisotropies from the \emph{Planck} satellite~\cite{Planck:2018vyg} have shifted the CMB-inferred value downwards to around $68$ km/s/Mpc with a tight uncertainty of $0.5$ km/s/Mpc (the exact value depends slightly on which datasets are used).

The possible systematics of either measurement have been scrutinized, without revealing any single systematic that could be responsible for this tension (see, e.g., Refs.~\cite{Riess:2021jrx,Abdalla:2022yfr}). More worryingly, multiple different probes now cluster in two `camps' of either high or low preferred value of $H_0$ \cite{Verde:2023lmm}. While direct kinematical probes of the recent expansion history (such as from the distance ladder calibrated on Cepheids, Miras, other variable stars, or the TRGB, as well as maser and strong lensing measurements) generally prefer a higher value of the Hubble constant, the indirect probes relying on a dynamical model of the universe (such as the CMB anisotropies, the combination of BAO and BBN, as well as more generally the clustering of the LSS) generally prefer lower values of the Hubble constant. This striking preference of a roughly $\sim 7\%$ larger $H_0$ value between different families of probes \cite{Verde:2023lmm} clearly motivates the need to look for physics beyond \LCDM.

A large number of such models have been proposed, see \cite{Schoneberg:2021qvd,DiValentino:2021izs,Khalife:2023qbu}, but only a few of them have been relatively successful, and none can claim that they unambiguously resolve the tension. Modifications to the post-recombination, `late-time', universe are disfavored due to the fact that once SNeIa are calibrated using any method in the distance ladder, their luminosity distance can be compared to that of the BAO calibrated using the usual sound horizon from the CMB, leading to a clear disagreement over the same range of redshifts \cite{Efstathiou:2021ocp,Raveri:2023zmr,Poulin:2024ken}. Indeed, late-time solutions generally struggle with the multitude of observations that prefer an expansion history close to that of the $\Lambda$CDM model \cite{Schoneberg:2021qvd}. This points towards models which reduce the value of the sound horizon inferred from the CMB, and therefore modify physics at or before recombination. 

These kinds of models can generally be grouped into three categories, based on the mechanism responsible for reducing the sound horizon. Typically, these models shorten the amount of physical time that the primordial sound waves can travel, leading to a smaller sound horizon. This can be accomplished either by (a) shifting the redshift of recombination to be earlier, (b) increasing the pre-recombination expansion rate through additional dark radiation, or (c) increasing the pre-recombination expansion rate through another contribution to the energy density. Successful models of type (a) include a time-dependent variation of the electron mass (e.g., Ref.~\cite{Hart:2019dxi,Schoneberg:2024ynd}) and, to a lesser extent, primordial magnetic fields \cite{Thiele:2021okz,Galli:2021mxk,Jedamzik:2023rfd}. Successful models of type (b) include self-interacting dark radiation \cite{Bashinsky:2003tk,lesgourgues2013neutrino,Baumann:2015rya,Schoneberg:2021qvd}, Wess-Zumino dark radiation (WZDR) \cite{Aloni:2021eaq,Schoneberg:2022grr,Schoneberg:2023rnx}, and a Majoron model \cite{Escudero:2019gvw,EscuderoAbenza:2020egd,Escudero:2021rfi}. Successful models of type (c) include early dark energy (EDE) \cite{Karwal:2016vyq,Poulin:2018cxd}, early modified gravity (EMG) \cite{Adi:2020qqf,Braglia:2020auw}, new (cold) early dark energy (NEDE) \cite{Niedermann:2019olb,Niedermann:2020dwg,Cruz:2023lmn}, and other similar proposals. 

If any of these models successfully address the Hubble tension, future observations must be able to clearly establish significant deviations from \LCDM. For example, it is well-known that additional free-streaming dark radiation causes a phase shift in the acoustic peaks of the CMB and increases the amount of diffusion damping leading to a suppression in power above multipoles of about $\ell \sim 2000$ \cite{Planck:2018vyg}. However, since any successful model must fit current CMB data, which is itself well-fit by \LCDM, we know that a (likely complex) set of parameter degeneracies must come into play which makes identifying simple signatures of any given model difficult. Axion-like EDE is a case in point: ACT DR4 showed a significant preference for this model, along with a significant increase in the CMB-inferred value of $H_0$, but it was unclear what exact features which drove this preference \cite{Hill:2021yec,Poulin:2021bjr,Smith:2022hwi}. As we show, the impact of most of these models on multipoles greater than $\ell \sim 2000$ -- at scales that are not as well measured by \emph{Planck} -- leads to clear and simple deviations from \LCDM. The ability to measure these deviations will depend on how well systematic errors from foreground modeling and the non-linear growth of matter perturbations are limited. 

We note that there are other approaches to resolving the Hubble tension which may not lead to any strong deviations from \LCDM predictions in the CMB. For example, a re-scaling of the gravitational free-fall rates and photon-electron scattering rate leaves most dimensionless cosmological observables nearly invariant, leaving the Hubble constant relatively unconstrained from CMB measurements alone \cite{Cyr-Racine:2021oal,Baryakhtar:2024rky}. Such a scenario may be realized by proposing a \enquote*{mirror world} dark sector. The initial scenario presented in Ref.~\cite{Cyr-Racine:2021oal} does not work since it transfers the Hubble tension to a tension in the primordial helium abundance but further model building (e.g., Ref.~\cite{Greene:2024qis}) may prove to be interesting. Of course these models make other \enquote*{beyond the standard model} predictions, which will allow them to be tested with future data. A related degeneracy has been discussed in the context of electron mass variations in \cite{Schoneberg:2024ynd}, which can be probed by large-scale structure data.

\enlargethispage*{3\baselineskip}
This work sets out to explore the possible observational consequences of some of the most successful models proposed to ease the Hubble tension. We describe the employed methodology in \cref{sec:method} and report the results for a few well motivated example models in \cref{sec:results}. Finally, we discuss our results in \cref{sec:discussion} and conclude in \cref{sec:conclusion}.

\section{Method}\label{sec:method}

In order to predict what kind of small-scale power spectra we can expect to observe in future CMB anisotropy observations for the considered models, we focus on the most probable model parameter space preferred by current-day CMB experiments. Despite this seemingly simple goal, there are several details to be aware of:
\begin{itemize}
    \item Since our goal is to establish clearly measurable predictions of solutions to the Hubble tension that are allowed by current data, we focus on the part of parameter space where these models are able to best address the Hubble tension. For this purpose, following Refs.~\cite{Benevento:2020fev,Camarena:2021jlr,Efstathiou:2021ocp,Schoneberg:2021qvd}, we impose a Gaussian prior on the Type Ia supernovae absolute magnitude, $M_b = -19.253 \pm 0.027$ \cite{Brout:2022vxf}.
    \item It is important to state that this is a conservative estimate, since it might be that a model can still be preferred as only a partial solution to the Hubble tension despite seemingly being \enquote*{excluded} by future data. In this case the CMB would exclude a full resolution of the Hubble tension but might still be compatible with a partial one. 
    \item Additionally, there may be shifts in other cosmological parameters (such as $\Omega_m$, $n_s$, or $\Omega_b h^2$) that are slightly disfavored by current data but favored by future data. This could either boost or reduce the constraining power for the models we investigate.
    \item We purposefully use the somewhat older BAO data (see e.g.~Ref.~\cite{Schoneberg:2021qvd}) that was used in the \emph{Planck} legacy analysis \cite{Planck:2018vyg} in order to have a more direct comparison with the corresponding data and best fit files. However, we do expect newer BAO data (such as from the recently released completed eBOSS survey \cite{eBOSS:2020yzd} as well as those from DESI \cite{DESI:2025zgx}) to cause small parameter shifts. Given the slight discrepancies between SDSS and DESI in extended parameter space (see, e.g., Ref.~\cite{Chaussidon:2025tww}), we decide to use older (and less constraining) BAO data. We use the Pantheon+\cite{Brout:2022vxf} SNeIa data, and note that the use of DES-Y5 \cite{DES:2024jxu} or Union-3 \cite{Rubin:2023ovl} would also result in small shifts to parameter values due to their preference for different values of $\Omega_m$\,.
    \item The impact of non-linear structure formation has to be crucially taken into account, as otherwise the analyses would be severely biased. We discuss our approach to this issue in \cref{app:nonlinear} (see also Ref.~\cite{Trendafilova:2025dce}).
    \item Importantly, for some of the considered models the changes in other parameters (such as $n_s$ and $\Omega_m$/$\Omega_m h^2$) might be more measurable using complementary data from galaxy clustering or weak lensing. We focus in this work only on the measurability through near-future CMB data.
    \item Our estimates of the sensitivity of near-future CMB measurements do not account for uncertainties due to modeling of foreground contamination. As such, the error bars present an idealized lower-limit. We also note that foreground modeling can introduce strong correlations between multipoles. 
\end{itemize}
In summary, our results present an unambiguous prediction for the shape of the angular power spectrum of the primary anisotropies from these models, but more work would need to be done to translate this into an clean estimate of the detectability of these models in future data.

\enlargethispage*{2\baselineskip}
For the purpose of this study, we use higher precision settings adjusted for near-future CMB data. In particular, these data are expected to measure much smaller angular scales and therefore are more sensitive to small-scale perturbations. This requires both higher accuracy in the underlying Boltzmann codes as well as a careful consideration of the non-linear prescription (for the latter see \cref{app:nonlinear}).

We use \texttt{class} as the underlying Einstein-Boltzmann code to predict the power spectra. We note that since \texttt{class} version $>3.2.3$ the in-built precision has been increased by the developers (using a new way of computing the lensing power spectrum based on the Limber approximation that becomes increasingly more accurate at smaller scales), but for older versions the precision needs to be manually increased by increasing the \texttt{k\_max\_tau0\_over\_lmax} precision parameter (in our case to $15$). In either case, despite predicting the lensed CMB power spectra only to $\ell = 5000$ it is crucial for accurate lensing predictions to predict the lensing power spectrum at least to $\ell = 8000$ (in our case using the parameters \texttt{l\_max\_scalars=7000}, and the default \texttt{delta\_l\_max\_lensing=1000}). We also set \texttt{accurate\_lensing=1} and \texttt{l\_switch\_limber=9}. Note that for the MCMC runs we use the low-precision (and hence quicker) default settings of these codes, only requiring high-precision settings when performing the predictions for near-future CMB observations.

For the MCMC runs we use the \texttt{MontePython} code~\cite{Audren:2012wb,Brinckmann:2018cvx}, interfaced with several CMB likelihoods that are already released or are to be released in the near future. Our core analysis uses the \texttt{plik} likelihood analyzing the PR3 data \cite{Planck:2019nip}. In order to explore if using \emph{Planck} PR4 data (NPIPE) \cite{Planck:2020olo} has an impact, we also analyzed our results using the \texttt{CamSpec} likelihood \cite{Rosenberg:2022sdy}. Though, as we discuss in \cref{app:likelihoods}, we find that this has a mostly negligible impact on the small-scale predictions, and therefore focus on the PR3 predictions here. We use the BAO data from BOSS DR12 \cite{BOSS:2016wmc}, MGS \cite{Ross:2014qpa}, and 6dFGS \cite{2011MNRAS.416.3017B} and the Pantheon+ SNeIa data from \cite{Brout:2022vxf}. We use the default priors for each likelihood, as well as flat unbounded priors for each of the cosmological parameters, enforcing only $\tau_\mathrm{reio}>0.004$. As in \cite{Planck:2018vyg} we use a single massive neutrino of $0.06\mathrm{eV}$ mass. The priors on the parameters which are specific to each of the non-\LCDM\ models are uninformative. 

\section{Results}\label{sec:results}

We investigate three main models in this work. These are chosen as the most successful models investigated in Ref.~\cite{Schoneberg:2021qvd} (including variations on their baseline mechanisms), following the three categories outlined in the Introduction. For each of the models we show the predicted power spectra with a prior on the Type Ia supernovae absolute magnitude, $M_b$ (equivalent to one on $H_0$), compared to the reference\footnote{The $\Lambda$CDM reference cosmology has $\Omega_b h^2=0.02242$, $\Omega_\mathrm{cdm}h^2=0.11933$, $H_0 = 67.66 \mathrm{km/s/Mpc}$, $\tau_\mathrm{reio}=0.0561$, $A_s = 2.105 \cdot 10^{-9}$, $n_s = 0.9665$, and a single massive neutrino of $0.06\mathrm{eV}$ mass, which are the officially reported mean values in \cite{Planck:2018vyg}.} $\Lambda$CDM power spectrum obtained without a prior on $M_b$\,. Explicitly, we define the difference $\Delta C_\ell^{\rm XY} \equiv (C_\ell^{\rm XY})_{{\rm new\ model}, M_b} - (C_\ell^{\rm XY})_{\Lambda{\rm CDM}, \mathrm{no}~M_b}$ for the supplied figures. We randomly draw 100 points from the part of the MCMC chain that has the burn-in removed in order to draw the mean and standard deviation lines/contours; we note that increasing this number to 1000 points did not significantly change the results.
 
The power spectra predicted by these samples are then compared to current-day uncertainties from the \emph{Planck} experiment (blue error bars) and an estimate of the uncertainties from ACT DR6 (gray error bars) and CMB-S4 (black error bars) as described in \cref{app:uncertainties}. We also show the uncertainties on the CMB anisotropies due to the non-linear structure formation treatment (and baryonic feedback) as discussed in \cref{app:nonlinear} as red (and purple) regions. 

\subsection{Wess-Zumino dark radiation}

As discussed before, in order to increase the CMB-inferred Hubble constant it is necessary to decrease the sound horizon \cite{Schoneberg:2021qvd,Poulin:2024ken}. The introduction of \enquote*{dark radiation} -- radiation which effectively does not interact with visible matter -- is a natural choice. There are many possible extensions of the core model of particle physics which can achieve this, chief among them an increase in the total energy density in neutrinos. However, such a change is significantly limited by the fact that neutrinos free-stream, leading to a measurable phase shift in the acoustic oscillations \cite{Bashinsky:2003tk,Follin:2015hya,Baumann:2019keh,Saravanan:2025cyi,Montefalcone:2025unv}. On the other hand, strongly self-interacting dark radiation models do not lead to changes in the phase as well as partially compensate the increased diffusion damping due to enhanced clustering on smaller scales (see \cite[Sec.~3.2]{Blinov:2020hmc}) and are currently a viable way to address the Hubble tension. However, as is the case with all models that introduce additional energy density in the pre-recombination universe, these models significantly impact the diffusion damping scale \mbox{\cite{Brust:2017nmv,Blinov:2020hmc,Becker:2020hzj,Archidiacono:2019wdp,Saravanan:2025cyi}}.

In addition to considering strongly self-interacting dark radiation, one can also change when its abundance is generated. Often times this is done either through decay of some heaver particle or through thermalization of the dark radiation bath with another heat bath. For example, in order to avoid constraints from measurements of the light element abundances and big bang nucleosynthesis (BBN), the dark radiation may be produced after BBN (e.g., Ref.~\cite{Aloni:2023tff}). The same mechanism can also be employed for the CMB itself. If the time during which the dark radiation abundance is increased is set at a redshift of around $z \gtrsim 10^4$ (about a decade in redshift before recombination), modes on the smaller scales ($\ell \gtrsim 1000$) that are very sensitive to diffusion damping will have experienced a smaller amount of dark radiation when entering the Hubble horizon, while modes on the larger scales ($\ell \lesssim 1000$) which are also those scales for which the sound horizon can be very well measured will have experienced a larger amount of dark radiation. Overall, this leads to the possibility of having a small sound horizon with a comparatively smaller amount of diffusion damping experienced on the smallest scales \cite{Schoneberg:2022grr}.

\begin{figure}
    \centering
    \includegraphics[width=1\linewidth]{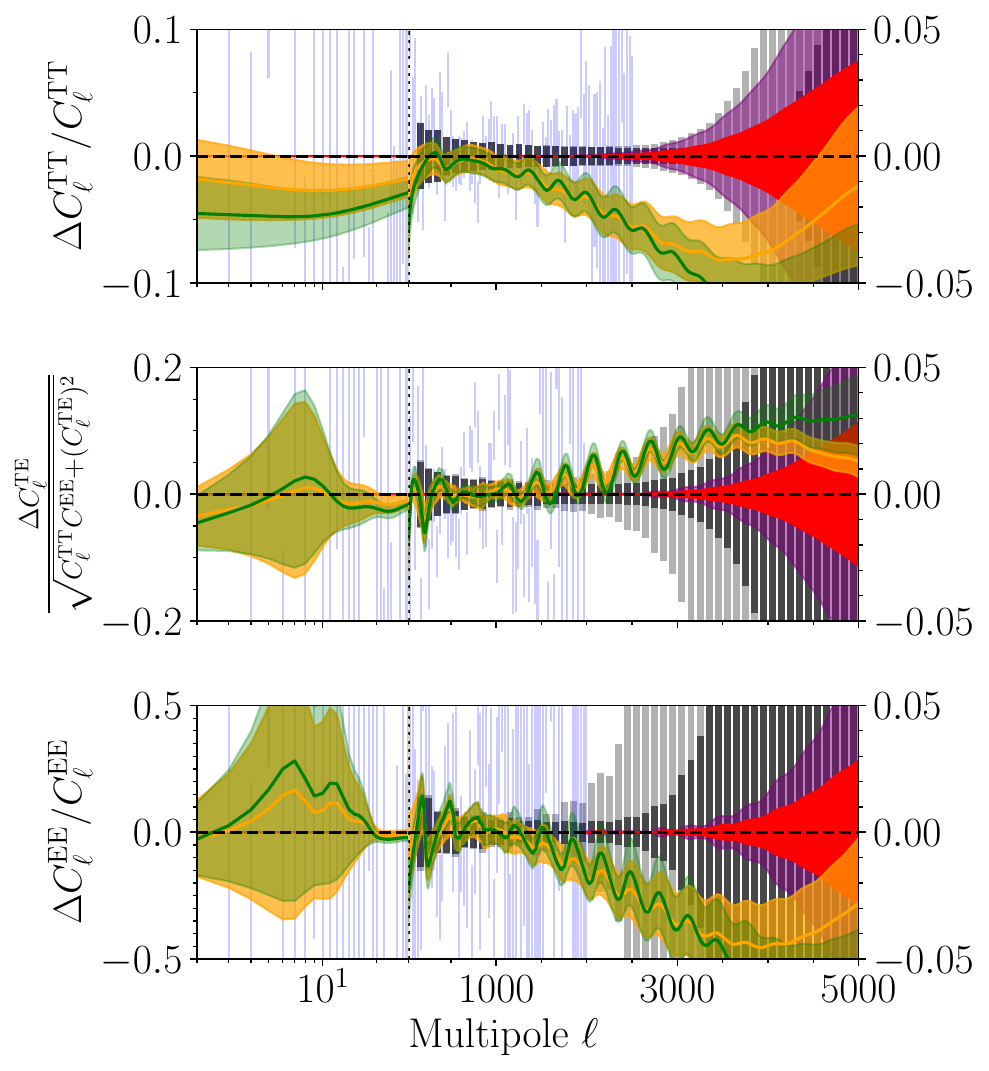}
    \caption{We show the difference between the predicted power spectra and a reference $\Lambda$CDM power spectrum for the TT, TE, and EE correlations, as described in the text. We compare it to the \emph{Planck} measurements in blue, the predicted uncertainties for a future ACT (CMB-S4) data release in gray (black), as well as uncertainties from the non-linear treatment in red/purple according to \cref{app:nonlinear}. We show the predictions for the baseline WZDR model in orange and for the model with additional interactions in green.}
    \label{fig:wzdr}
\end{figure}

This kind of model has been constructed using Wess-Zumino dark radiation (WZDR) in \cite{Aloni:2021eaq,Schoneberg:2022grr}. We use the implementation outlined in \cite{Schoneberg:2022grr,Schoneberg:2023rnx}. The predicted small-scale power spectra are shown in \cref{fig:wzdr}. We show a model without any interactions in orange and a model where the dark radiation interacts with the dark matter (as in \cite{Joseph:2022jsf,Schoneberg:2023rnx}) in green.

It is evident that this kind of model may be detected/excluded by near-future CMB data, especially when no significant interaction between the dark radiation and dark matter is present. We comment on the upturn in TT and EE, for the non-interacting model, at $\ell \gtrsim 3000$ in \cref{sec:discussion}.

\subsection{Early dark energy}\label{ssec:ede}

Similar to dark radiation models, the EDE mechanism introduces additional energy density to the pre-recombination universe. However, unlike the dark radiation models, EDE models introduce a new cosmological field which makes a significant contribution to the total energy density over a relatively limited amount of time. See Ref.~\cite{Poulin:2023lkg} for a review.

\enlargethispage*{1\baselineskip}
In the initial proposal in Ref.~\cite{Poulin:2018cxd} the model proposes the existence of a single scalar field in an axion-like potential. In that model the field is initially frozen at a non-zero value of its potential due to Hubble friction, serving effectively as dark energy until it becomes dynamical. Once the Hubble friction becomes less than the acceleration from the local potential slope the field begins to oscillate, quickly diluting away its energy density. This effectively increases the expansion ratio around and especially before the recombination but otherwise leaves the later universe unchanged. See Refs.~\cite{Simon:2022adh,Smith:2023oop,Efstathiou:2023fbn} for recent constraints. 

These models are typically specified by four quantities: the redshift at which the EDE energy density makes its maximum fractional contribution to the total energy density, the maximum amplitude of this fractional contribution, the rate at which the EDE energy density redshifts after this peak, and a sound speed which governs how perturbations evolve. 

We have computed the predictions from a scalar field EDE model following Refs.~\cite{Poulin:2018cxd, Smith:2019ihp} using a scalar field potential which scales as $\phi^6$ around its minimum, thereby fixing the rate at which the EDE energy density redshifts after its peak to be $\rho_\phi \propto a^{-4.5}$ \cite{Turner:1983he}. In addition to this we have computed the predictions using a fluid parameterization \cite{Simon:2023hlp,Lin:2019qug,Poulin:2023lkg} in order to explore whether there are aspects of the small-scale CMB power spectrum in these models that are unique to the scalar field dynamics.  In the fluid case we have left the equation of state after the peak EDE contribution as an additional free parameter. 

\begin{figure}
    \centering
    \includegraphics[width=1\linewidth]{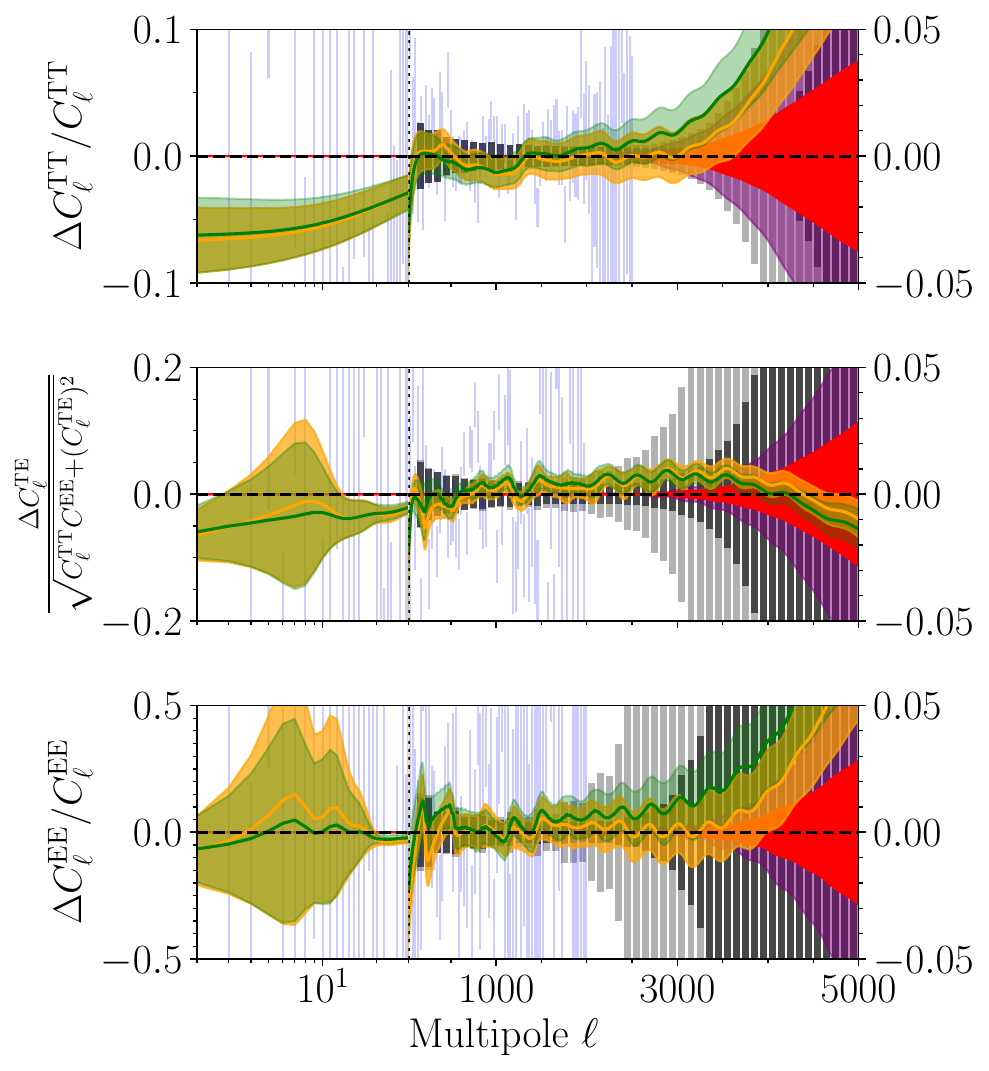}
    \caption{Same as \cref{fig:wzdr} but for EDE (orange) fluid-like EDE model (green).}
    \label{fig:ede}
\end{figure}

The constraints for both these variations are shown in \cref{fig:ede}. The fluid approximation deviates slightly more than the axion-like early dark energy model. However, generally the two models behave very similarly. We do observe that they deviate especially in the temperature autocorrelation on the smallest scales (at $\ell \gtrsim 3000$), but in a way that is mostly compatible with the expected uncertainty of ACT DR6 and somewhat degenerate with the non-linear (baryonic feedback) treatment. 

We have also investigated whether additional freedom of varying $Y_\mathrm{He}$ (which impacts the small-scale damping of the CMB as $\theta_{\rm d} \propto (1-Y_{\rm He}/2)^{-1/2}$) leads a significant change to the upturn at $\ell \gtrsim 3000$. With current CMB data \cite{Planck:2018vyg} we find the allowed range of $Y_{\rm He} = 0.241^{+0.016}_{-0.011}$ is too small to make a noticeable impact. 

Since the initial proposal of EDE, a number of different variations have been investigated. In \cite{Niedermann:2019olb,Niedermann:2020dwg,Cruz:2023lmn,Chatrchyan:2024xjj}
the authors investigate a (cold) \enquote{new early dark energy} (NEDE) model where the dynamics of a scalar field is precipitated by a first order phase transition in the dark sector which is connected to the dynamics of a second scalar field, called the `trigger' field. The original proposal in \cite{Niedermann:2019olb,Niedermann:2020dwg} assumed that the trigger field energy density is negligible, but it has been shown in \cite{Cruz:2023lmn,Chatrchyan:2024xjj} (here tNEDE) that abandoning this assumption gives additional rich phenomenology, such as reduced clustering on the smallest scales.

The original NEDE model is specified by three parameters: the redshift of the phase transition, the maximum fractional energy density contained within the scalar fields, and the equation of state which describes how this energy density redshifts as the phase transition progresses; the effective sound speed for the perturbations is taken to be equal to this equation of state. The tNEDE model introduces the energy density in the trigger field today as an additional parameter. 

\begin{figure}
    \centering
    \includegraphics[width=1\linewidth]{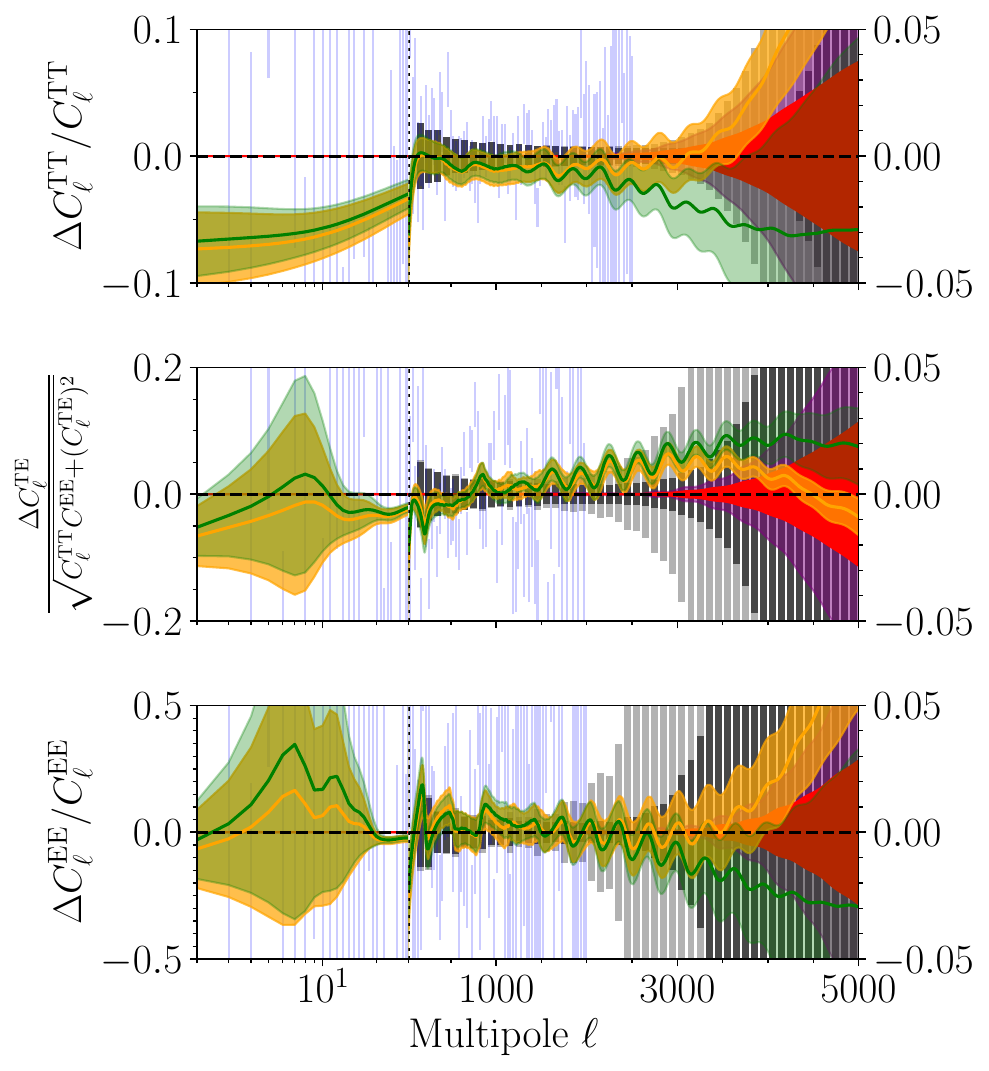}
    \caption{Same as \cref{fig:wzdr} but for cold NEDE (orange) and a variation with non-negligible trigger field energy density (green), tNEDE.}
    \label{fig:nede}
\end{figure}

We show the constraints for this model in \cref{fig:nede}. For the original cold NEDE model the constraints are almost identical to that of the axion-like EDE, except for a clear upturn in the TE spectra around $\ell \gtrsim 1500$ that could be detectable with near-future data. On the other hand, tNEDE clearly shows a very different phenomenology on the smallest scales, with a non-zero posterior for the energy density of the trigger field today, $\Omega_{0,t} = 0.0042^{+0.0018}_{-0.0030}$ and a mass $m_t = 2.16^{+0.45}_{-0.58}\times 10^{-27}\ {\rm eV}$. In this case the model is generally more consistent with $\Lambda$CDM due to the additional suppression of the small-scale power-- $S_8 =0.818^{+0.022}_{-0.019}$ in tNEDE versus $S_8 = 0.847\pm 0.013$ in NEDE.  The two models show similar features in the TE spectrum around $\ell \sim 1000$ that might be detectable. We note that while the mean value in the temperature autocorrelation for this variation is much more discrepant from the $\Lambda$CDM reference due to the additional suppression of small-scale power, the $1 \sigma$ contour is quite broad, making it somewhat uncertain how well this feature may be measured with future CMB data. 

In the absence of an ultra-light scalar field the predicted small-scale power spectra for all EDE models considered here are very similar, showing a clear upturn in both TT and EE starting around $\ell \gtrsim 3000$. As we argue in \cref{sec:discussion}, such a feature is generic for any model that addresses the Hubble tension by injecting a significant amount of energy density over a limited amount of time. 

\subsection{Varying electron mass}
\label{sec:var_me}

There are several indications that our current standard model of particle physics is not the final description of the universe (lacking explanations for dark energy and dark matter, a neutrino mass mechanism, as well as a quantum theory of gravity) and it is natural to ask if fundamental \enquote*{constants} are allowed to vary in time. In particular, an electron mass, $m_e$, which was slightly larger around recombination than it is today has proven to be surprisingly effective at resolving the cosmic conundrum of the Hubble tension as well as being slightly favored by recent DESI data, see \cite{Hart:2019dxi,Sekiguchi:2020teg,Schoneberg:2021qvd,Schoneberg:2024ynd,Baryakhtar:2024rky,Baryakhtar:2025uxs}. With a larger value of $m_e$ the redshift of recombination is increased, leading to a smaller sound horizon and therefore a larger allowed Hubble constant while keeping the CMB angular sound horizon scale unchanged. Indeed, in \cite{Schoneberg:2024ynd,Sekiguchi:2020teg,Baryakhtar:2024rky} it has been argued that the CMB might be entirely insensitive to such a change at first order if other cosmological parameters are varied concurrently. See \cite{Hart:2017ndk,Chiang:2018xpn,Hart:2021kad,Chluba:2023xqj,Hart:2019dxi,Sekiguchi:2020teg,Lynch:2024hzh,Baryakhtar:2025uxs} for further constraints, and \cite{Schoneberg:2024ynd} for a review.

In particular, the small-scale CMB is left nearly invariant if the relative abundance of baryons and photons, matter and radiation, and the angular size of the sound horizon at recombination is fixed \cite{Sekiguchi:2020teg,Baryakhtar:2024rky}. The resulting scaling gives, $h \propto m_e^{3.18}$ and $\Omega_m h^2 \propto m_e$ \cite{Baryakhtar:2024rky}. The first scaling law tells us that to resolve the Hubble tension we need $m^{\rm early}_e/m^{\rm late}_e \simeq 1.02$ which, in turn, \emph{limits} the physical matter density to increase by at most $\sim 2\%$.  Large-scale CMB measurements and late-time probes of the expansion history (such as SNeIa and BAO) give tight constraints on $\Omega_m$, which, along with the required increase in $h$ (and, in turn, $\Omega_m h^2$), leads to limitations on this model's ability to fully address the Hubble tension (even more complicated changes to the recombination history face the same constraints \cite{Lee:2022gzh})\footnote{As shown in the next section, dark radiation and EDE models are less subject to this since they introduce additional degeneracies allowing for $\Omega_m h^2$ to reach larger values.}. One way to mitigate this is to include additional degrees of freedom for the late-time expansion history, such as allowing for non-zero curvature, as we do below. 

Here we implement the model as described in \cite{Schoneberg:2024ynd} using a phenomenological approach -- noting that a microphysical theory of electron mass variations is described in Ref.~\cite{Baryakhtar:2024rky} (see also Ref.~\cite{Schoneberg:2024ynd}) and constrained, using a wide range of data sets, in Ref.~\cite{Baryakhtar:2025uxs}. In these theories the mass of the electron is controlled by an ultra-light scalar field which becomes dynamical some time between recombination and today. 

We describe the model using a single additional parameter $m_e^\mathrm{early}/m_e^\mathrm{late}$ describing the relative difference of the electron mass around CMB times compared to that measured in the laboratory today. We assume a transition in the dark ages where electromagnetic observational data is sparse, see also Ref.~\cite{Schoneberg:2024ynd} for more details on other observational probes for this model, including a discussion on why BBN light element abundances are currently not constraining enough to restrict such variations.

\begin{figure}
    \centering
    \includegraphics[width=1\linewidth]{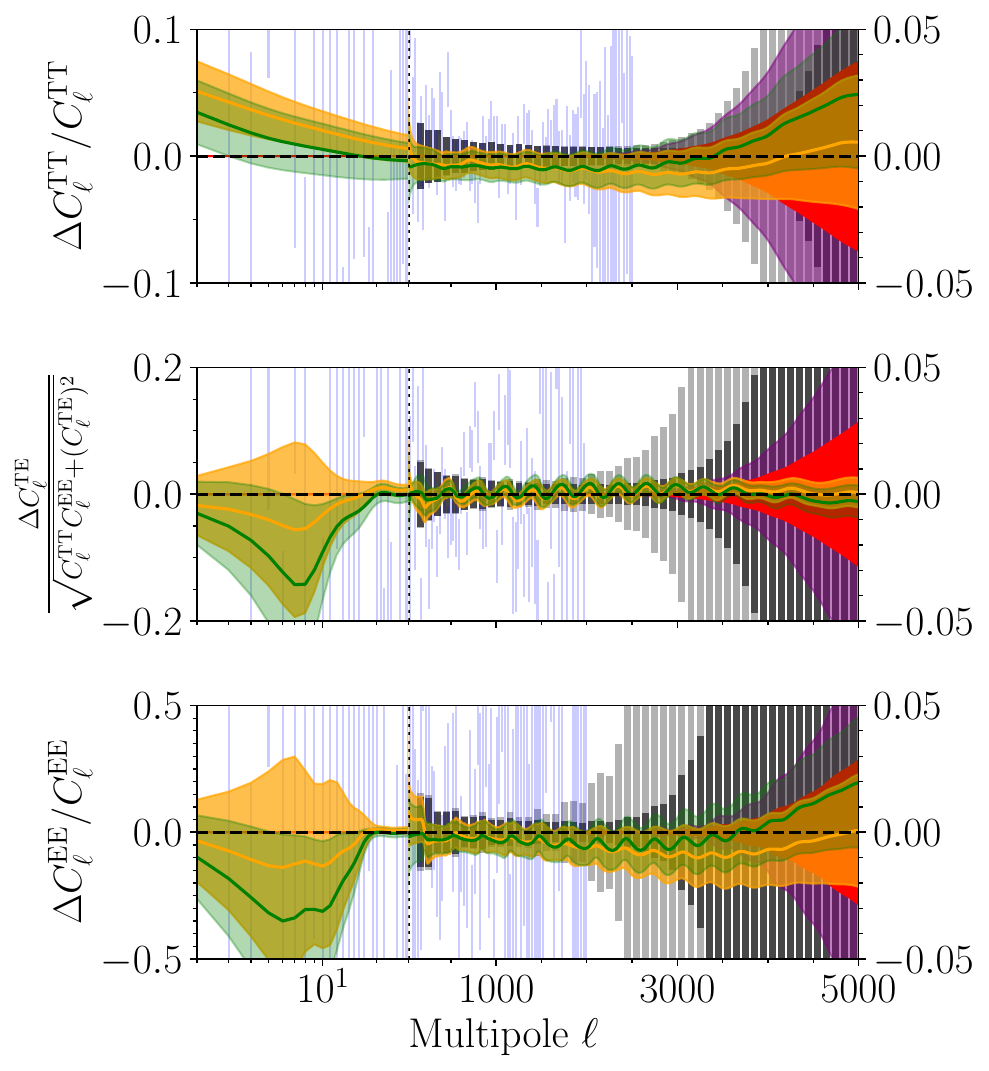}
    \caption{As in \cref{fig:wzdr}, but for a model with a time-variation of the electron mass. We show the predictions for the baseline model in orange and with additional curvature allowed in green.}
    \label{fig:me}
\end{figure}

\enlargethispage*{1\baselineskip}
We show the predicted power spectra in \cref{fig:me}. We observe that the varying $m_e$ model can leave the intermediate to small-scale CMB invariant with respect to \LCDM\ and is therefore mostly indistinguishable from \LCDM\ despite having a much larger value of $H_0$ (around $H_0 = 71.2\pm 0.7$km/s/Mpc with $m_e^\mathrm{early}/m_e^\mathrm{late} = 1.0198 \pm 0.0044$, a roughly 2\% increase in the electron mass at CMB times) \cite{Sekiguchi:2020teg,Schoneberg:2024ynd,Baryakhtar:2024rky}. A a slightly large $H_0$ can be reached when including curvature (green curve and colored region in \cref{fig:me}). In this case there are larger oscillations in the mean prediction, but overall the result is not far outside the expected uncertainty bands for near-future CMB surveys and the mean value is still mostly compatible with zero. In this case we find $H_0 = 71.8 \pm 1.1$km/s/Mpc with $m_e^\mathrm{early}/m_e^\mathrm{late} = 1.037 \pm 0.013$, a roughly 4\% increase in the electron mass at CMB times, and $\Omega_k = (-6.5 \pm  4.5) \cdot 10^{-3}$.

\section{Discussion}
\label{sec:discussion}

\begin{figure}
    \centering
    \includegraphics[width=1\linewidth]{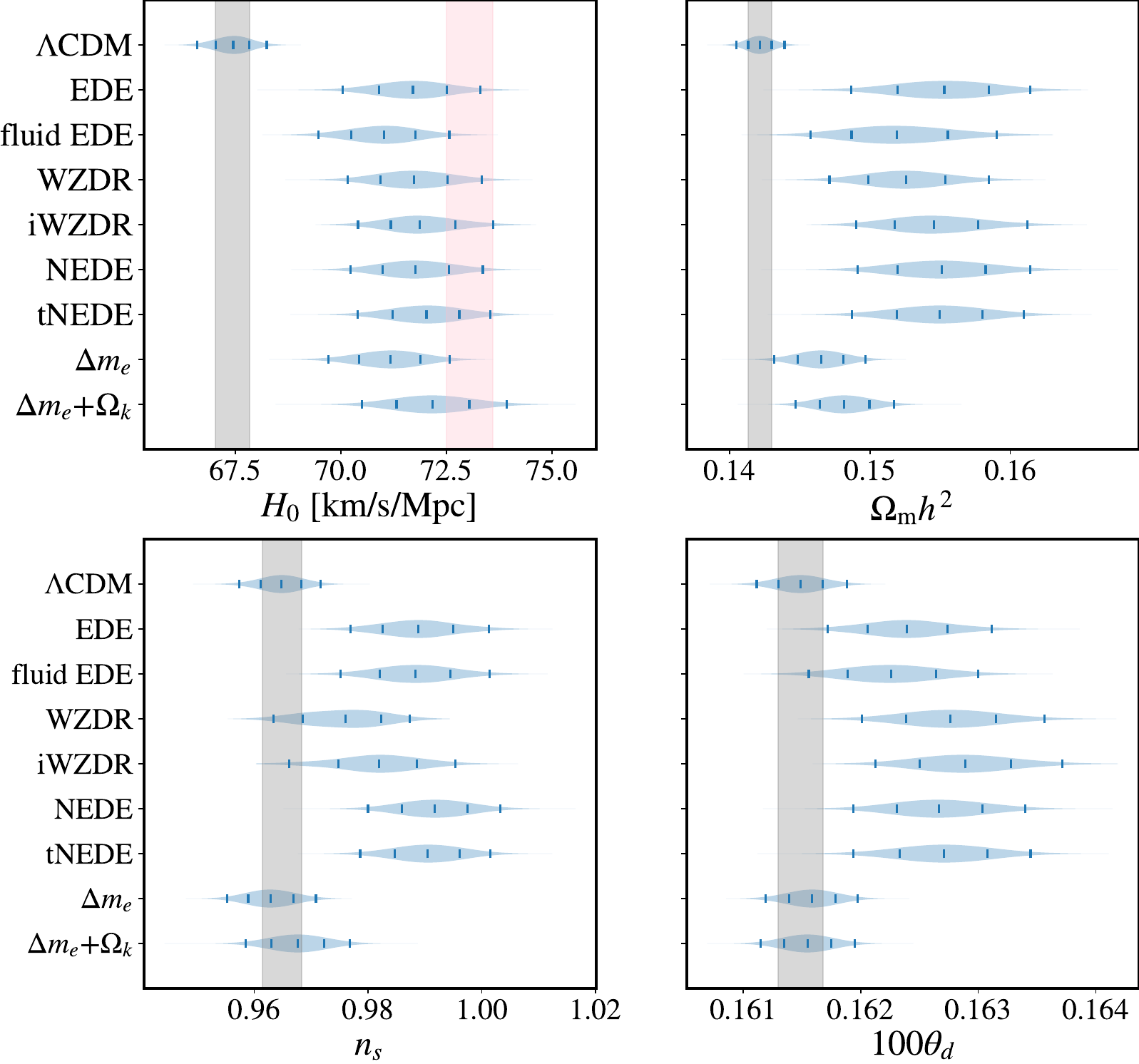}
    \caption{Violin-plot of the distributions of a few selected cosmological parameters for the models investigated in this work. The gray vertical bands denote the 1$\sigma$ contours from \emph{Planck}, while the blue bars show the mean, 1$\sigma$, and $2\sigma$ constraints for the respective parameter. The pink vertical band in the $H_0$ plot denotes the 1$\sigma$ constrain from the SH0ES collaboration \cite{Brout:2022vxf}.}
    \label{fig:whisker}
\end{figure}

We show the constraints for a selected number of cosmological parameters for all models investigated in this work in \cref{fig:whisker}. All of the models achieve a relatively high value of $H_0$ (around 71-72 km/s/Mpc). Most of the models accomplish the high $H_0$ while increasing $n_s$\,, $\Omega_\mathrm{m}h^2$ and $\theta_{\rm d}$ -- except for those that shift the recombination redshift ($\Delta m_e$/$\Delta m_e + \Omega_k$). 

As was argued in Ref.~\cite{Poulin:2023lkg}, models which are described by \LCDM in the late universe but increase the pre-recombination expansion rate leading to an increase in the CMB-inferred value of of the Hubble constant to $fH_0$ will invariably increase the diffusion damping scale, $\theta_{\rm d}$, by a factor $\sim f^{1/2}$. In addition to this, in order to keep measurements of $\Omega_m$ using BAO and SNeIa fixed, the physical matter density, $\Omega_m h^2$, must increase by a factor of $\sim f^2$. All of these trends are broadly confirmed by the parameter shifts seen in \cref{fig:whisker}.  

The \emph{increase} in the diffusion damping scale for the \enquote*{pure} EDE models (excluding tNEDE due to the addition of a non-clustering matter component) would seem to imply that the small-scale CMB temperature and polarization power spectra for these models should show a decrement at high multipoles compared to \LCDM. However our results show that there is instead an \emph{enhancement} of power in these models. This enhancement is caused by the increased matter clustering in these models due to an increase of both $n_s$ and $\Omega_m h^2$ (and to a lesser extent $A_s$ \cite{Smith:2020rxx}), which leads to an increase in the matter power spectrum on the smaller scales, leading to stronger lensing of the CMB. Indeed the power from lensing becomes an important part of the lensed CMB anisotropies at $\ell \gtrsim 3000$ and starts to dominate at scales $\ell \gg 3000$, as shown for example in \cite[Fig.~6]{Lewis:2006fu}.

\begin{figure}
    \centering
    \includegraphics[width=1\linewidth]{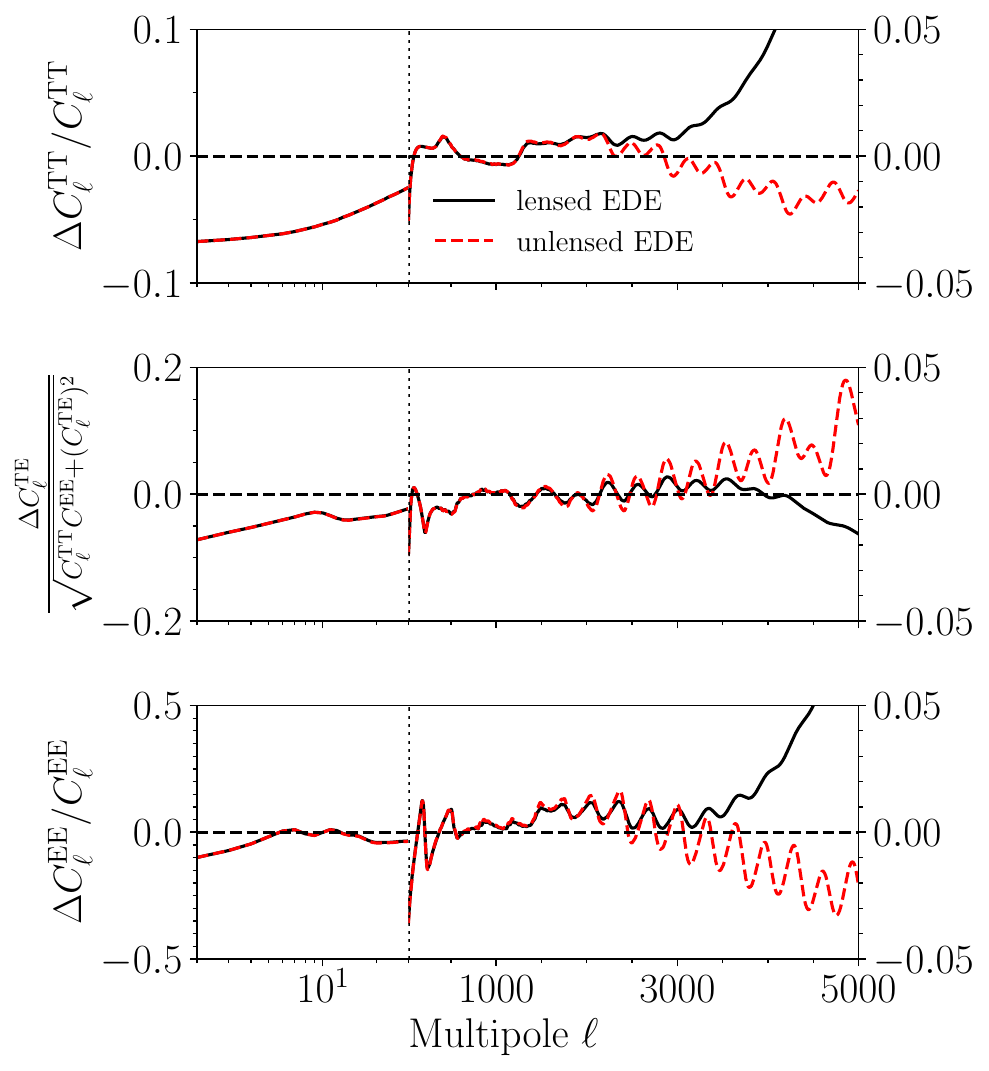}
    \caption{A comparison between the residual of the best fit axion-like EDE model with and without lensing. We can see that the upturn is due to the lensing contribution. Note that when computing the unlensed residual we used a \LCDM\ reference power spectrum which is also unlensed.}
    \label{fig:EDE_lens_nolens}
\end{figure}

We compare the lensed and unlensed residuals for the best fit axion-like EDE model in \cref{fig:EDE_lens_nolens}. There we can see that the unlensed residuals remain relatively flat compared to the dramatic upturn once lensing is included. 

Given that WZDR also introduces additional pre-recombination energy density and increases $\theta_{\rm d}$ we might expect it to show a similar behavior. However, as seen in \cref{fig:wzdr} we instead see a decrease in power. This difference requires additional explanation. 

\begin{figure}
    \centering
    \includegraphics[width=1\linewidth]{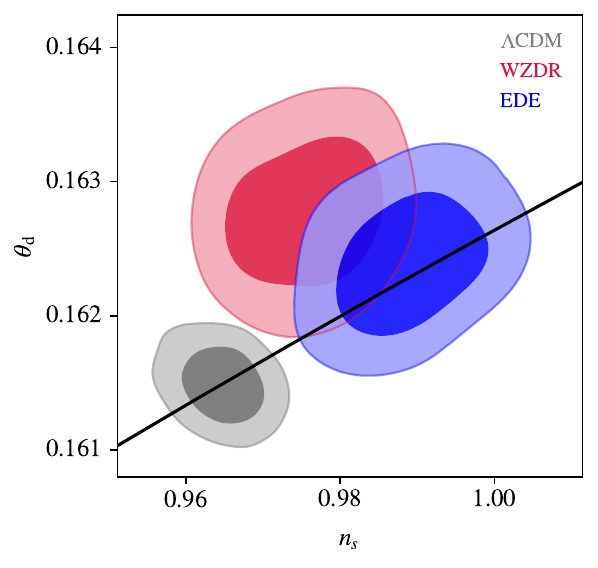}
    \caption{The black line shows the degeneracy which leaves the residual small-scale CMB relatively unchanged, given in Eq.~\ref{eq:ns_thetad}. It is clear that WZDR lies significantly above this degeneracy (which corresponds to a decrement compared to \LCDM), whereas the axion-like EDE posteriors lie along it.}
    \label{fig:theta_d_vs_ns}
\end{figure}

One part of the explanation is due to the lack of balance between the small-scale effect of $\theta_{\rm d}$ and $n_s$. As was shown in Ref.~\cite[III.A]{Smith:2023oop}, both parameters need to increase to keep the small-scale CMB relatively unchanged (compared to the \LCDM mean values). Their Eq.~(7) reads
\begin{equation}
    \delta \theta_{\rm d}/\theta_{\rm d} = 0.2 \delta n_s/n_s. \label{eq:ns_thetad}
\end{equation}
\Cref{fig:theta_d_vs_ns} shows the posteriors of $\theta_{\rm d}$ and $n_s$ for \LCDM, WZDR, and axion-like EDE. The black line shows the degeneracy given in \cref{eq:ns_thetad} with $\delta \theta_{\rm d} \equiv \theta_{\rm d}-\bar{\theta}^{\Lambda{\rm CDM}}_{\rm d}$ and $\delta n_s  \equiv n_s-\bar{n}^{\Lambda{\rm CDM}}_{s}$. It is clear that the axion-like EDE model lies along this degeneracy whereas WZDR is significantly above it. Models which are above the line have a damping scale which is \emph{larger}, leading to enhanced damping on small scales. 

In addition to this, in WZDR the lensing has a smaller amplitude than in EDE models. The overall amplitude of the lensing potential power spectrum scales with $A_s k_{\rm eq} \chi_{\rm dec}$ \cite{Planck:2015mym,Smith:2022iax}, where $k_{\rm eq}$ is the wavenumber entering the Hubble horizon at matter/radiation equality and $\chi_{\rm dec}$ is the comoving distance to the surface of last scattering. Since $\Omega_m$ is relatively fixed between the models and $H_0$ cancels out in the expression, the only leftover dependence is the amount of radiation, which determines $k_\mathrm{eq}$. In WZDR models the amount of radiation is increased, leading to later matter-radiation equality, smaller $k_\mathrm{eq}$\,, and correspondingly less lensing of the CMB compared to a pure EDE model.

Variations in the electron mass leave both the damping scale and $n_s$ unchanged as well as have tighter upper limits on $\Omega_m h^2$ due to the degeneracies introduced by these models between $m_e$, $H_0$ and $\Omega_m h^2$, as discussed in section \ref{sec:var_me}. Additionally, as argued in \cite{Schoneberg:2024ynd, Baryakhtar:2024rky} there is a strong degeneracy for all underlying scales (damping, angular, equality, ...) involved in the CMB that should cause this model to be very close to \LCDM, which is what we observe.

\section{Conclusions}\label{sec:conclusion}

With the advent of precision cosmology the Hubble tension has become a serious problem for the standard $\Lambda$CDM model. In particular, since the arcminute-scale CMB anisotropies have been observed by the \emph{Planck} satellite this tension has grown drastically. In this work we have explored the possible observational consequences of some of the most successful models that have been proposed to ease the Hubble tension, focusing specifically on new physics in the CMB damping tail.

We find that some, but not all, of these models make predictions for clear deviations from \LCDM in the small-scale CMB power spectrum. In particular, we find that WZDR (with or without interactions) predicts a decrement in power on small scales, with an upturn at even smaller scales. Pure EDE models all show features in the TT and TE spectra around $\ell \sim 1000$ that might be borderline observable, as well as an overall upturn of the TT and EE power spectra at $\ell \sim 3000$. When a non-clustering dark matter component is added to EDE, as in the case of NEDE with a non-negligible trigger field energy density, the small scale power may be suppressed at $\ell \sim 3000$. Finally, we find that variations in the electron mass leads to a relatively flat fractional change in power. 

\begin{figure}
    \centering
    \includegraphics[width=1\linewidth]{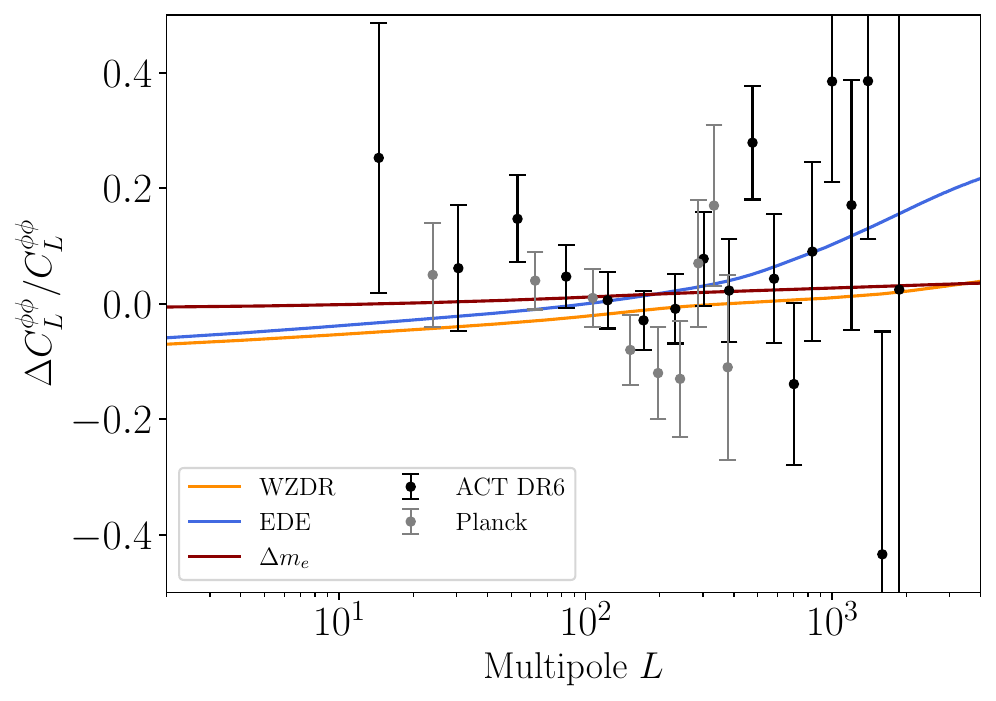}
    \caption{The lensing potential power spectrum predictions from the best fit predictions of three of the models we consider here. Note that, as discussed in \cref{sec:discussion} the EDE prediction rises above the others at large multipoles.}
    \label{fig:lensing}
\end{figure}

Far future measurements of the lensing potential power spectrum may also play a role in distinguishing these models. We show a comparison between the best fit predictions of these models and our baseline \LCDM model in \cref{fig:lensing}. Limitations to measuring the lensing potential power spectrum at large angular scales likely prevents us from detecting a deficit of power there. However, the enhanced power in EDE models may be detected with far future CMB measurements, with the caveat that the way we model non-linear growth of structure will have a significant effect on these predictions. 

It is also interesting to consider how `de-lensing' may play a role in detecting these deviations \cite{2012JCAP...06..014S,Hotinli:2021umk}. However, we note that in the limit of reconstructing the true primary CMB some of the key features -- such as the enhancement of power in pure EDE models -- are removed (see \cref{fig:EDE_lens_nolens}). We leave an exploration of how de-lensing may affect our conclusions to future work. 

Near-future CMB data are expected to be able to potentially detect the damping tail predictions of some, but not all, of the investigated solutions to the Hubble tension. The predicted deviations for WZDR are significantly larger than the expected error bars of near-future CMB experiments, indicating that its potential as a viable solution may be clearly decided relatively soon with the upcoming ACT DR6 and SPT datasets. On the other hand, EDE models (with or without a non-clustering matter component) predict deviations from \LCDM at such small scales that one needs to be concerned about the accuracy of the modeling of the non-linear growth of structure as well as baryonic feedback. Once that modeling is under control it may be possible to detect these features with CMB observatories such as CMB-S4 and the Simons Observatory. Finally, variations in the electron mass make predictions for the small-scale CMB which may remain indistinguishable from \LCDM even with future CMB observations in line with the analytical arguments of \cite{Sekiguchi:2020teg,Schoneberg:2024ynd,Baryakhtar:2024rky}. 

In addition to having to develop new techniques to model the non-linear growth of structure, any future detection of these features also requires proper modeling of foreground contamination. For example, the upturn in the fractional power predicted by pure EDE models is at the several percent level -- an effect that may be degenerate with terms in the foreground modeling. 

This is not to say that improved CMB data cannot put pressure on these models. For example, improved measurements of the intermediate scale CMB ($\ell \sim 1000-2000$) have already put pressure on some of the models considered here. The most up-to-date analysis of the \emph{Planck} satellite data (NPIPE) increased its precision relative to its original data release \cite{Planck:2019nip} (plik), leading to improved constraints on both pure EDE and NEDE models \cite{Efstathiou:2023fbn,Chatrchyan:2024xjj}, with $H_0 = 71.81_{-0.87}^{+0.88}$ km/s/Mpc using plik becoming $H_0 = 71.25_{-0.81}^{+0.8}$ km/s/Mpc with NPIPE for the pure EDE model. On the other hand, constraints on the WZDR and varying electron mass models are only minimally changed with the updated \emph{Planck} data: $H_0 = 71.76_{-0.82}^{+0.82}$ km/s/Mpc with plik becoming $H_0 = 71.83_{-0.83}^{+0.82}$ km/s/Mpc with NPIPE for WZDR and $H_0 =72.18_{-0.9}^{+0.86}$ km/s/Mpc with plik going to $H_0 = 72.27_{-0.88}^{+0.88}$ km/s/Mpc with NPIPE for a varying electron mass with non-zero curvature. All of these constraints use the data sets described in section \ref{sec:method}.

It is important to note that unique predictions of these models appear in many other cosmological data sets. Shifts in the cosmological parameters for many of these models are likely to be observable in future late-time data such as from the large scale structure probed by galaxy surveys or from the luminosity distance of SNeIa. 

Our work makes it clear that near-future measurements of the CMB damping tail may rule in or out whole classes of models of the early universe which attempt to address the Hubble tension. What is particularly exciting is that these models make predictions for features in the CMB which are clearly distinguishable from \LCDM, and from one another, and therefore may provide unambiguous evidence for new physics in the pre-recombination universe. 

\textbf{Note} While this manuscript was being finalized the results of ACT DR6 were announced \cite{Louis:2025tst,Calabrese:2025mza}; the main results were produced before the announcement. As anticipated by our analysis, ACT DR6 is unable to place significant constraints on EDE (see Sec.~5.2.1 and Fig.~12 of Ref.~\cite{Calabrese:2025mza}), and in fact shows a slight preference for EDE with an improvement in the $\chi^2$ of up to 6.6 (see Table 2). Similarly, Sec.~5.2.1 shows that when fit to ACT DR6, a varying electron mass plus curvature is able to raise the CMB-inferred value of $H_0$. 

\begin{acknowledgements}

The authors thank Silvia Galli, Vivian Poulin, Daniel Grin, Zachary J.~Weiner, and Ali Rida Khalife for useful conversations. TLS is supported by NSF Grants No.~2009377 and No.~2308173. Computational work for this was, in part, performed on Firebird, a cluster supported by Swarthmore College and Lafayette College. NS gratefully acknowledges the support by a Fraunhofer-Schwarzschild Fellowship at Universitäts-Sternwarte München.

\end{acknowledgements}

\bibliography{biblio.bib}

\begin{thebibliography}{102}%
\makeatletter
\providecommand \@ifxundefined [1]{%
 \@ifx{#1\undefined}
}%
\providecommand \@ifnum [1]{%
 \ifnum #1\expandafter \@firstoftwo
 \else \expandafter \@secondoftwo
 \fi
}%
\providecommand \@ifx [1]{%
 \ifx #1\expandafter \@firstoftwo
 \else \expandafter \@secondoftwo
 \fi
}%
\providecommand \natexlab [1]{#1}%
\providecommand \enquote  [1]{``#1''}%
\providecommand \bibnamefont  [1]{#1}%
\providecommand \bibfnamefont [1]{#1}%
\providecommand \citenamefont [1]{#1}%
\providecommand \href@noop [0]{\@secondoftwo}%
\providecommand \href [0]{\begingroup \@sanitize@url \@href}%
\providecommand \@href[1]{\@@startlink{#1}\@@href}%
\providecommand \@@href[1]{\endgroup#1\@@endlink}%
\providecommand \@sanitize@url [0]{\catcode `\\12\catcode `\$12\catcode
  `\&12\catcode `\#12\catcode `\^12\catcode `\_12\catcode `\%12\relax}%
\providecommand \@@startlink[1]{}%
\providecommand \@@endlink[0]{}%
\providecommand \url  [0]{\begingroup\@sanitize@url \@url }%
\providecommand \@url [1]{\endgroup\@href {#1}{\urlprefix }}%
\providecommand \urlprefix  [0]{URL }%
\providecommand \Eprint [0]{\href }%
\providecommand \doibase [0]{http://dx.doi.org/}%
\providecommand \selectlanguage [0]{\@gobble}%
\providecommand \bibinfo  [0]{\@secondoftwo}%
\providecommand \bibfield  [0]{\@secondoftwo}%
\providecommand \translation [1]{[#1]}%
\providecommand \BibitemOpen [0]{}%
\providecommand \bibitemStop [0]{}%
\providecommand \bibitemNoStop [0]{.\EOS\space}%
\providecommand \EOS [0]{\spacefactor3000\relax}%
\providecommand \BibitemShut  [1]{\csname bibitem#1\endcsname}%
\let\auto@bib@innerbib\@empty
\bibitem [{\citenamefont {{Peebles}}(2020)}]{2020RvMP...92c0501P}%
  \BibitemOpen
  \bibfield  {author} {\bibinfo {author} {\bibfnamefont {P.~J.~E.}\
  \bibnamefont {{Peebles}}},\ }\href {\doibase 10.1103/RevModPhys.92.030501}
  {\bibfield  {journal} {\bibinfo  {journal} {Reviews of Modern Physics}\
  }\textbf {\bibinfo {volume} {92}},\ \bibinfo {eid} {030501} (\bibinfo {year}
  {2020})}\BibitemShut {NoStop}%
\bibitem [{\citenamefont {Brout}\ \emph {et~al.}(2022)\citenamefont {Brout}
  \emph {et~al.}}]{Brout:2022vxf}%
  \BibitemOpen
  \bibfield  {author} {\bibinfo {author} {\bibfnamefont {D.}~\bibnamefont
  {Brout}} \emph {et~al.},\ }\href {\doibase 10.3847/1538-4357/ac8e04}
  {\bibfield  {journal} {\bibinfo  {journal} {Astrophys. J.}\ }\textbf
  {\bibinfo {volume} {938}},\ \bibinfo {pages} {110} (\bibinfo {year}
  {2022})},\ \Eprint {http://arxiv.org/abs/2202.04077} {arXiv:2202.04077
  [astro-ph.CO]} \BibitemShut {NoStop}%
\bibitem [{\citenamefont {Riess}\ \emph {et~al.}(2022)\citenamefont {Riess}
  \emph {et~al.}}]{Riess:2021jrx}%
  \BibitemOpen
  \bibfield  {author} {\bibinfo {author} {\bibfnamefont {A.~G.}\ \bibnamefont
  {Riess}} \emph {et~al.},\ }\href {\doibase 10.3847/2041-8213/ac5c5b}
  {\bibfield  {journal} {\bibinfo  {journal} {Astrophys. J. Lett.}\ }\textbf
  {\bibinfo {volume} {934}},\ \bibinfo {pages} {L7} (\bibinfo {year} {2022})},\
  \Eprint {http://arxiv.org/abs/2112.04510} {arXiv:2112.04510 [astro-ph.CO]}
  \BibitemShut {NoStop}%
\bibitem [{\citenamefont {Freedman}(2021)}]{Freedman:2021ahq}%
  \BibitemOpen
  \bibfield  {author} {\bibinfo {author} {\bibfnamefont {W.~L.}\ \bibnamefont
  {Freedman}},\ }\href {\doibase 10.3847/1538-4357/ac0e95} {\bibfield
  {journal} {\bibinfo  {journal} {Astrophys. J.}\ }\textbf {\bibinfo {volume}
  {919}},\ \bibinfo {pages} {16} (\bibinfo {year} {2021})},\ \Eprint
  {http://arxiv.org/abs/2106.15656} {arXiv:2106.15656 [astro-ph.CO]}
  \BibitemShut {NoStop}%
\bibitem [{\citenamefont {Aghanim}\ \emph
  {et~al.}(2020{\natexlab{a}})\citenamefont {Aghanim} \emph
  {et~al.}}]{Planck:2018vyg}%
  \BibitemOpen
  \bibfield  {author} {\bibinfo {author} {\bibfnamefont {N.}~\bibnamefont
  {Aghanim}} \emph {et~al.} (\bibinfo {collaboration} {Planck}),\ }\href
  {\doibase 10.1051/0004-6361/201833910} {\bibfield  {journal} {\bibinfo
  {journal} {Astron. Astrophys.}\ }\textbf {\bibinfo {volume} {641}},\ \bibinfo
  {pages} {A6} (\bibinfo {year} {2020}{\natexlab{a}})},\ \bibinfo {note}
  {[Erratum: Astron.Astrophys. 652, C4 (2021)]},\ \Eprint
  {http://arxiv.org/abs/1807.06209} {arXiv:1807.06209 [astro-ph.CO]}
  \BibitemShut {NoStop}%
\bibitem [{\citenamefont {Abdalla}\ \emph {et~al.}(2022)\citenamefont {Abdalla}
  \emph {et~al.}}]{Abdalla:2022yfr}%
  \BibitemOpen
  \bibfield  {author} {\bibinfo {author} {\bibfnamefont {E.}~\bibnamefont
  {Abdalla}} \emph {et~al.},\ }\href {\doibase 10.1016/j.jheap.2022.04.002}
  {\bibfield  {journal} {\bibinfo  {journal} {JHEAp}\ }\textbf {\bibinfo
  {volume} {34}},\ \bibinfo {pages} {49} (\bibinfo {year} {2022})},\ \Eprint
  {http://arxiv.org/abs/2203.06142} {arXiv:2203.06142 [astro-ph.CO]}
  \BibitemShut {NoStop}%
\bibitem [{\citenamefont {Verde}\ \emph {et~al.}(2024)\citenamefont {Verde},
  \citenamefont {Sch\"oneberg},\ and\ \citenamefont
  {Gil-Mar\'\i{}n}}]{Verde:2023lmm}%
  \BibitemOpen
  \bibfield  {author} {\bibinfo {author} {\bibfnamefont {L.}~\bibnamefont
  {Verde}}, \bibinfo {author} {\bibfnamefont {N.}~\bibnamefont {Sch\"oneberg}},
  \ and\ \bibinfo {author} {\bibfnamefont {H.}~\bibnamefont {Gil-Mar\'\i{}n}},\
  }\href {\doibase 10.1146/annurev-astro-052622-033813} {\bibfield  {journal}
  {\bibinfo  {journal} {Ann. Rev. Astron. Astrophys.}\ }\textbf {\bibinfo
  {volume} {62}},\ \bibinfo {pages} {287} (\bibinfo {year} {2024})},\ \Eprint
  {http://arxiv.org/abs/2311.13305} {arXiv:2311.13305 [astro-ph.CO]}
  \BibitemShut {NoStop}%
\bibitem [{\citenamefont {Sch\"oneberg}\ \emph {et~al.}(2022)\citenamefont
  {Sch\"oneberg}, \citenamefont {Franco~Abell\'an}, \citenamefont
  {P\'erez~S\'anchez}, \citenamefont {Witte}, \citenamefont {Poulin},\ and\
  \citenamefont {Lesgourgues}}]{Schoneberg:2021qvd}%
  \BibitemOpen
  \bibfield  {author} {\bibinfo {author} {\bibfnamefont {N.}~\bibnamefont
  {Sch\"oneberg}}, \bibinfo {author} {\bibfnamefont {G.}~\bibnamefont
  {Franco~Abell\'an}}, \bibinfo {author} {\bibfnamefont {A.}~\bibnamefont
  {P\'erez~S\'anchez}}, \bibinfo {author} {\bibfnamefont {S.~J.}\ \bibnamefont
  {Witte}}, \bibinfo {author} {\bibfnamefont {V.}~\bibnamefont {Poulin}}, \
  and\ \bibinfo {author} {\bibfnamefont {J.}~\bibnamefont {Lesgourgues}},\
  }\href {\doibase 10.1016/j.physrep.2022.07.001} {\bibfield  {journal}
  {\bibinfo  {journal} {Phys. Rept.}\ }\textbf {\bibinfo {volume} {984}},\
  \bibinfo {pages} {1} (\bibinfo {year} {2022})},\ \Eprint
  {http://arxiv.org/abs/2107.10291} {arXiv:2107.10291 [astro-ph.CO]}
  \BibitemShut {NoStop}%
\bibitem [{\citenamefont {Di~Valentino}\ \emph {et~al.}(2021)\citenamefont
  {Di~Valentino}, \citenamefont {Mena}, \citenamefont {Pan}, \citenamefont
  {Visinelli}, \citenamefont {Yang}, \citenamefont {Melchiorri}, \citenamefont
  {Mota}, \citenamefont {Riess},\ and\ \citenamefont
  {Silk}}]{DiValentino:2021izs}%
  \BibitemOpen
  \bibfield  {author} {\bibinfo {author} {\bibfnamefont {E.}~\bibnamefont
  {Di~Valentino}}, \bibinfo {author} {\bibfnamefont {O.}~\bibnamefont {Mena}},
  \bibinfo {author} {\bibfnamefont {S.}~\bibnamefont {Pan}}, \bibinfo {author}
  {\bibfnamefont {L.}~\bibnamefont {Visinelli}}, \bibinfo {author}
  {\bibfnamefont {W.}~\bibnamefont {Yang}}, \bibinfo {author} {\bibfnamefont
  {A.}~\bibnamefont {Melchiorri}}, \bibinfo {author} {\bibfnamefont {D.~F.}\
  \bibnamefont {Mota}}, \bibinfo {author} {\bibfnamefont {A.~G.}\ \bibnamefont
  {Riess}}, \ and\ \bibinfo {author} {\bibfnamefont {J.}~\bibnamefont {Silk}},\
  }\href {\doibase 10.1088/1361-6382/ac086d} {\bibfield  {journal} {\bibinfo
  {journal} {Class. Quant. Grav.}\ }\textbf {\bibinfo {volume} {38}},\ \bibinfo
  {pages} {153001} (\bibinfo {year} {2021})},\ \Eprint
  {http://arxiv.org/abs/2103.01183} {arXiv:2103.01183 [astro-ph.CO]}
  \BibitemShut {NoStop}%
\bibitem [{\citenamefont {Khalife}\ \emph {et~al.}(2024)\citenamefont
  {Khalife}, \citenamefont {Zanjani}, \citenamefont {Galli}, \citenamefont
  {G\"unther}, \citenamefont {Lesgourgues},\ and\ \citenamefont
  {Benabed}}]{Khalife:2023qbu}%
  \BibitemOpen
  \bibfield  {author} {\bibinfo {author} {\bibfnamefont {A.~R.}\ \bibnamefont
  {Khalife}}, \bibinfo {author} {\bibfnamefont {M.~B.}\ \bibnamefont
  {Zanjani}}, \bibinfo {author} {\bibfnamefont {S.}~\bibnamefont {Galli}},
  \bibinfo {author} {\bibfnamefont {S.}~\bibnamefont {G\"unther}}, \bibinfo
  {author} {\bibfnamefont {J.}~\bibnamefont {Lesgourgues}}, \ and\ \bibinfo
  {author} {\bibfnamefont {K.}~\bibnamefont {Benabed}},\ }\href {\doibase
  10.1088/1475-7516/2024/04/059} {\bibfield  {journal} {\bibinfo  {journal}
  {JCAP}\ }\textbf {\bibinfo {volume} {04}},\ \bibinfo {pages} {059} (\bibinfo
  {year} {2024})},\ \Eprint {http://arxiv.org/abs/2312.09814} {arXiv:2312.09814
  [astro-ph.CO]} \BibitemShut {NoStop}%
\bibitem [{\citenamefont {Efstathiou}(2021)}]{Efstathiou:2021ocp}%
  \BibitemOpen
  \bibfield  {author} {\bibinfo {author} {\bibfnamefont {G.}~\bibnamefont
  {Efstathiou}},\ }\href {\doibase 10.1093/mnras/stab1588} {\bibfield
  {journal} {\bibinfo  {journal} {Mon. Not. Roy. Astron. Soc.}\ }\textbf
  {\bibinfo {volume} {505}},\ \bibinfo {pages} {3866} (\bibinfo {year}
  {2021})},\ \Eprint {http://arxiv.org/abs/2103.08723} {arXiv:2103.08723
  [astro-ph.CO]} \BibitemShut {NoStop}%
\bibitem [{\citenamefont {Raveri}(2023)}]{Raveri:2023zmr}%
  \BibitemOpen
  \bibfield  {author} {\bibinfo {author} {\bibfnamefont {M.}~\bibnamefont
  {Raveri}},\ }\href@noop {} {\  (\bibinfo {year} {2023})},\ \Eprint
  {http://arxiv.org/abs/2309.06795} {arXiv:2309.06795 [astro-ph.CO]}
  \BibitemShut {NoStop}%
\bibitem [{\citenamefont {Poulin}\ \emph {et~al.}(2024)\citenamefont {Poulin},
  \citenamefont {Smith}, \citenamefont {Calder\'on},\ and\ \citenamefont
  {Simon}}]{Poulin:2024ken}%
  \BibitemOpen
  \bibfield  {author} {\bibinfo {author} {\bibfnamefont {V.}~\bibnamefont
  {Poulin}}, \bibinfo {author} {\bibfnamefont {T.~L.}\ \bibnamefont {Smith}},
  \bibinfo {author} {\bibfnamefont {R.}~\bibnamefont {Calder\'on}}, \ and\
  \bibinfo {author} {\bibfnamefont {T.}~\bibnamefont {Simon}},\ }\href@noop {}
  {\  (\bibinfo {year} {2024})},\ \Eprint {http://arxiv.org/abs/2407.18292}
  {arXiv:2407.18292 [astro-ph.CO]} \BibitemShut {NoStop}%
\bibitem [{\citenamefont {Hart}\ and\ \citenamefont
  {Chluba}(2020)}]{Hart:2019dxi}%
  \BibitemOpen
  \bibfield  {author} {\bibinfo {author} {\bibfnamefont {L.}~\bibnamefont
  {Hart}}\ and\ \bibinfo {author} {\bibfnamefont {J.}~\bibnamefont {Chluba}},\
  }\href {\doibase 10.1093/mnras/staa412} {\bibfield  {journal} {\bibinfo
  {journal} {Mon. Not. Roy. Astron. Soc.}\ }\textbf {\bibinfo {volume} {493}},\
  \bibinfo {pages} {3255} (\bibinfo {year} {2020})},\ \Eprint
  {http://arxiv.org/abs/1912.03986} {arXiv:1912.03986 [astro-ph.CO]}
  \BibitemShut {NoStop}%
\bibitem [{\citenamefont {Sch\"oneberg}\ and\ \citenamefont
  {Vacher}(2025)}]{Schoneberg:2024ynd}%
  \BibitemOpen
  \bibfield  {author} {\bibinfo {author} {\bibfnamefont {N.}~\bibnamefont
  {Sch\"oneberg}}\ and\ \bibinfo {author} {\bibfnamefont {L.}~\bibnamefont
  {Vacher}},\ }\href {\doibase 10.1088/1475-7516/2025/03/004} {\bibfield
  {journal} {\bibinfo  {journal} {JCAP}\ }\textbf {\bibinfo {volume} {03}},\
  \bibinfo {pages} {004} (\bibinfo {year} {2025})},\ \Eprint
  {http://arxiv.org/abs/2407.16845} {arXiv:2407.16845 [astro-ph.CO]}
  \BibitemShut {NoStop}%
\bibitem [{\citenamefont {Thiele}\ \emph {et~al.}(2021)\citenamefont {Thiele},
  \citenamefont {Guan}, \citenamefont {Hill}, \citenamefont {Kosowsky},\ and\
  \citenamefont {Spergel}}]{Thiele:2021okz}%
  \BibitemOpen
  \bibfield  {author} {\bibinfo {author} {\bibfnamefont {L.}~\bibnamefont
  {Thiele}}, \bibinfo {author} {\bibfnamefont {Y.}~\bibnamefont {Guan}},
  \bibinfo {author} {\bibfnamefont {J.~C.}\ \bibnamefont {Hill}}, \bibinfo
  {author} {\bibfnamefont {A.}~\bibnamefont {Kosowsky}}, \ and\ \bibinfo
  {author} {\bibfnamefont {D.~N.}\ \bibnamefont {Spergel}},\ }\href {\doibase
  10.1103/PhysRevD.104.063535} {\bibfield  {journal} {\bibinfo  {journal}
  {Phys. Rev. D}\ }\textbf {\bibinfo {volume} {104}},\ \bibinfo {pages}
  {063535} (\bibinfo {year} {2021})},\ \Eprint
  {http://arxiv.org/abs/2105.03003} {arXiv:2105.03003 [astro-ph.CO]}
  \BibitemShut {NoStop}%
\bibitem [{\citenamefont {Galli}\ \emph {et~al.}(2022)\citenamefont {Galli},
  \citenamefont {Pogosian}, \citenamefont {Jedamzik},\ and\ \citenamefont
  {Balkenhol}}]{Galli:2021mxk}%
  \BibitemOpen
  \bibfield  {author} {\bibinfo {author} {\bibfnamefont {S.}~\bibnamefont
  {Galli}}, \bibinfo {author} {\bibfnamefont {L.}~\bibnamefont {Pogosian}},
  \bibinfo {author} {\bibfnamefont {K.}~\bibnamefont {Jedamzik}}, \ and\
  \bibinfo {author} {\bibfnamefont {L.}~\bibnamefont {Balkenhol}},\ }\href
  {\doibase 10.1103/PhysRevD.105.023513} {\bibfield  {journal} {\bibinfo
  {journal} {Phys. Rev. D}\ }\textbf {\bibinfo {volume} {105}},\ \bibinfo
  {pages} {023513} (\bibinfo {year} {2022})},\ \Eprint
  {http://arxiv.org/abs/2109.03816} {arXiv:2109.03816 [astro-ph.CO]}
  \BibitemShut {NoStop}%
\bibitem [{\citenamefont {Jedamzik}\ \emph {et~al.}(2025)\citenamefont
  {Jedamzik}, \citenamefont {Abel},\ and\ \citenamefont
  {Ali-Haimoud}}]{Jedamzik:2023rfd}%
  \BibitemOpen
  \bibfield  {author} {\bibinfo {author} {\bibfnamefont {K.}~\bibnamefont
  {Jedamzik}}, \bibinfo {author} {\bibfnamefont {T.}~\bibnamefont {Abel}}, \
  and\ \bibinfo {author} {\bibfnamefont {Y.}~\bibnamefont {Ali-Haimoud}},\
  }\href {\doibase 10.1088/1475-7516/2025/03/012} {\bibfield  {journal}
  {\bibinfo  {journal} {JCAP}\ }\textbf {\bibinfo {volume} {03}},\ \bibinfo
  {pages} {012} (\bibinfo {year} {2025})},\ \Eprint
  {http://arxiv.org/abs/2312.11448} {arXiv:2312.11448 [astro-ph.CO]}
  \BibitemShut {NoStop}%
\bibitem [{\citenamefont {Bashinsky}\ and\ \citenamefont
  {Seljak}(2004)}]{Bashinsky:2003tk}%
  \BibitemOpen
  \bibfield  {author} {\bibinfo {author} {\bibfnamefont {S.}~\bibnamefont
  {Bashinsky}}\ and\ \bibinfo {author} {\bibfnamefont {U.}~\bibnamefont
  {Seljak}},\ }\href {\doibase 10.1103/PhysRevD.69.083002} {\bibfield
  {journal} {\bibinfo  {journal} {Phys. Rev. D}\ }\textbf {\bibinfo {volume}
  {69}},\ \bibinfo {pages} {083002} (\bibinfo {year} {2004})},\ \Eprint
  {http://arxiv.org/abs/astro-ph/0310198} {arXiv:astro-ph/0310198} \BibitemShut
  {NoStop}%
\bibitem [{\citenamefont {Lesgourgues}\ \emph {et~al.}(2013)\citenamefont
  {Lesgourgues}, \citenamefont {Mangano}, \citenamefont {Miele},\ and\
  \citenamefont {Pastor}}]{lesgourgues2013neutrino}%
  \BibitemOpen
  \bibfield  {author} {\bibinfo {author} {\bibfnamefont {J.}~\bibnamefont
  {Lesgourgues}}, \bibinfo {author} {\bibfnamefont {G.}~\bibnamefont
  {Mangano}}, \bibinfo {author} {\bibfnamefont {G.}~\bibnamefont {Miele}}, \
  and\ \bibinfo {author} {\bibfnamefont {S.}~\bibnamefont {Pastor}},\
  }\href@noop {} {\emph {\bibinfo {title} {Neutrino cosmology}}}\ (\bibinfo
  {publisher} {Cambridge University Press},\ \bibinfo {year}
  {2013})\BibitemShut {NoStop}%
\bibitem [{\citenamefont {Baumann}\ \emph {et~al.}(2016)\citenamefont
  {Baumann}, \citenamefont {Green}, \citenamefont {Meyers},\ and\ \citenamefont
  {Wallisch}}]{Baumann:2015rya}%
  \BibitemOpen
  \bibfield  {author} {\bibinfo {author} {\bibfnamefont {D.}~\bibnamefont
  {Baumann}}, \bibinfo {author} {\bibfnamefont {D.}~\bibnamefont {Green}},
  \bibinfo {author} {\bibfnamefont {J.}~\bibnamefont {Meyers}}, \ and\ \bibinfo
  {author} {\bibfnamefont {B.}~\bibnamefont {Wallisch}},\ }\href {\doibase
  10.1088/1475-7516/2016/01/007} {\bibfield  {journal} {\bibinfo  {journal}
  {JCAP}\ }\textbf {\bibinfo {volume} {01}},\ \bibinfo {pages} {007} (\bibinfo
  {year} {2016})},\ \Eprint {http://arxiv.org/abs/1508.06342} {arXiv:1508.06342
  [astro-ph.CO]} \BibitemShut {NoStop}%
\bibitem [{\citenamefont {Aloni}\ \emph {et~al.}(2022)\citenamefont {Aloni},
  \citenamefont {Berlin}, \citenamefont {Joseph}, \citenamefont {Schmaltz},\
  and\ \citenamefont {Weiner}}]{Aloni:2021eaq}%
  \BibitemOpen
  \bibfield  {author} {\bibinfo {author} {\bibfnamefont {D.}~\bibnamefont
  {Aloni}}, \bibinfo {author} {\bibfnamefont {A.}~\bibnamefont {Berlin}},
  \bibinfo {author} {\bibfnamefont {M.}~\bibnamefont {Joseph}}, \bibinfo
  {author} {\bibfnamefont {M.}~\bibnamefont {Schmaltz}}, \ and\ \bibinfo
  {author} {\bibfnamefont {N.}~\bibnamefont {Weiner}},\ }\href {\doibase
  10.1103/PhysRevD.105.123516} {\bibfield  {journal} {\bibinfo  {journal}
  {Phys. Rev. D}\ }\textbf {\bibinfo {volume} {105}},\ \bibinfo {pages}
  {123516} (\bibinfo {year} {2022})},\ \Eprint
  {http://arxiv.org/abs/2111.00014} {arXiv:2111.00014 [astro-ph.CO]}
  \BibitemShut {NoStop}%
\bibitem [{\citenamefont {Sch\"oneberg}\ and\ \citenamefont
  {Franco~Abell\'an}(2022)}]{Schoneberg:2022grr}%
  \BibitemOpen
  \bibfield  {author} {\bibinfo {author} {\bibfnamefont {N.}~\bibnamefont
  {Sch\"oneberg}}\ and\ \bibinfo {author} {\bibfnamefont {G.}~\bibnamefont
  {Franco~Abell\'an}},\ }\href {\doibase 10.1088/1475-7516/2022/12/001}
  {\bibfield  {journal} {\bibinfo  {journal} {JCAP}\ }\textbf {\bibinfo
  {volume} {12}},\ \bibinfo {pages} {001} (\bibinfo {year} {2022})},\ \Eprint
  {http://arxiv.org/abs/2206.11276} {arXiv:2206.11276 [astro-ph.CO]}
  \BibitemShut {NoStop}%
\bibitem [{\citenamefont {Sch\"oneberg}\ \emph {et~al.}(2023)\citenamefont
  {Sch\"oneberg}, \citenamefont {Franco~Abell\'an}, \citenamefont {Simon},
  \citenamefont {Bartlett}, \citenamefont {Patel},\ and\ \citenamefont
  {Smith}}]{Schoneberg:2023rnx}%
  \BibitemOpen
  \bibfield  {author} {\bibinfo {author} {\bibfnamefont {N.}~\bibnamefont
  {Sch\"oneberg}}, \bibinfo {author} {\bibfnamefont {G.}~\bibnamefont
  {Franco~Abell\'an}}, \bibinfo {author} {\bibfnamefont {T.}~\bibnamefont
  {Simon}}, \bibinfo {author} {\bibfnamefont {A.}~\bibnamefont {Bartlett}},
  \bibinfo {author} {\bibfnamefont {Y.}~\bibnamefont {Patel}}, \ and\ \bibinfo
  {author} {\bibfnamefont {T.~L.}\ \bibnamefont {Smith}},\ }\href {\doibase
  10.1103/PhysRevD.108.123513} {\bibfield  {journal} {\bibinfo  {journal}
  {Phys. Rev. D}\ }\textbf {\bibinfo {volume} {108}},\ \bibinfo {pages}
  {123513} (\bibinfo {year} {2023})},\ \Eprint
  {http://arxiv.org/abs/2306.12469} {arXiv:2306.12469 [astro-ph.CO]}
  \BibitemShut {NoStop}%
\bibitem [{\citenamefont {Escudero}\ and\ \citenamefont
  {Witte}(2020)}]{Escudero:2019gvw}%
  \BibitemOpen
  \bibfield  {author} {\bibinfo {author} {\bibfnamefont {M.}~\bibnamefont
  {Escudero}}\ and\ \bibinfo {author} {\bibfnamefont {S.~J.}\ \bibnamefont
  {Witte}},\ }\href {\doibase 10.1140/epjc/s10052-020-7854-5} {\bibfield
  {journal} {\bibinfo  {journal} {Eur. Phys. J. C}\ }\textbf {\bibinfo {volume}
  {80}},\ \bibinfo {pages} {294} (\bibinfo {year} {2020})},\ \Eprint
  {http://arxiv.org/abs/1909.04044} {arXiv:1909.04044 [astro-ph.CO]}
  \BibitemShut {NoStop}%
\bibitem [{\citenamefont {Escudero~Abenza}\ and\ \citenamefont
  {Witte}(2020)}]{EscuderoAbenza:2020egd}%
  \BibitemOpen
  \bibfield  {author} {\bibinfo {author} {\bibfnamefont {M.}~\bibnamefont
  {Escudero~Abenza}}\ and\ \bibinfo {author} {\bibfnamefont {S.~J.}\
  \bibnamefont {Witte}},\ }in\ \href@noop {} {\emph {\bibinfo {booktitle}
  {{Prospects in Neutrino Physics}}}}\ (\bibinfo {year} {2020})\ \Eprint
  {http://arxiv.org/abs/2004.01470} {arXiv:2004.01470 [hep-ph]} \BibitemShut
  {NoStop}%
\bibitem [{\citenamefont {Escudero}\ and\ \citenamefont
  {Witte}(2021)}]{Escudero:2021rfi}%
  \BibitemOpen
  \bibfield  {author} {\bibinfo {author} {\bibfnamefont {M.}~\bibnamefont
  {Escudero}}\ and\ \bibinfo {author} {\bibfnamefont {S.~J.}\ \bibnamefont
  {Witte}},\ }\href {\doibase 10.1140/epjc/s10052-021-09276-5} {\bibfield
  {journal} {\bibinfo  {journal} {Eur. Phys. J. C}\ }\textbf {\bibinfo {volume}
  {81}},\ \bibinfo {pages} {515} (\bibinfo {year} {2021})},\ \Eprint
  {http://arxiv.org/abs/2103.03249} {arXiv:2103.03249 [hep-ph]} \BibitemShut
  {NoStop}%
\bibitem [{\citenamefont {Karwal}\ and\ \citenamefont
  {Kamionkowski}(2016)}]{Karwal:2016vyq}%
  \BibitemOpen
  \bibfield  {author} {\bibinfo {author} {\bibfnamefont {T.}~\bibnamefont
  {Karwal}}\ and\ \bibinfo {author} {\bibfnamefont {M.}~\bibnamefont
  {Kamionkowski}},\ }\href {\doibase 10.1103/PhysRevD.94.103523} {\bibfield
  {journal} {\bibinfo  {journal} {Phys. Rev. D}\ }\textbf {\bibinfo {volume}
  {94}},\ \bibinfo {pages} {103523} (\bibinfo {year} {2016})},\ \Eprint
  {http://arxiv.org/abs/1608.01309} {arXiv:1608.01309 [astro-ph.CO]}
  \BibitemShut {NoStop}%
\bibitem [{\citenamefont {Poulin}\ \emph {et~al.}(2019)\citenamefont {Poulin},
  \citenamefont {Smith}, \citenamefont {Karwal},\ and\ \citenamefont
  {Kamionkowski}}]{Poulin:2018cxd}%
  \BibitemOpen
  \bibfield  {author} {\bibinfo {author} {\bibfnamefont {V.}~\bibnamefont
  {Poulin}}, \bibinfo {author} {\bibfnamefont {T.~L.}\ \bibnamefont {Smith}},
  \bibinfo {author} {\bibfnamefont {T.}~\bibnamefont {Karwal}}, \ and\ \bibinfo
  {author} {\bibfnamefont {M.}~\bibnamefont {Kamionkowski}},\ }\href {\doibase
  10.1103/PhysRevLett.122.221301} {\bibfield  {journal} {\bibinfo  {journal}
  {Phys. Rev. Lett.}\ }\textbf {\bibinfo {volume} {122}},\ \bibinfo {pages}
  {221301} (\bibinfo {year} {2019})},\ \Eprint
  {http://arxiv.org/abs/1811.04083} {arXiv:1811.04083 [astro-ph.CO]}
  \BibitemShut {NoStop}%
\bibitem [{\citenamefont {Adi}\ and\ \citenamefont
  {Kovetz}(2021)}]{Adi:2020qqf}%
  \BibitemOpen
  \bibfield  {author} {\bibinfo {author} {\bibfnamefont {T.}~\bibnamefont
  {Adi}}\ and\ \bibinfo {author} {\bibfnamefont {E.~D.}\ \bibnamefont
  {Kovetz}},\ }\href {\doibase 10.1103/PhysRevD.103.023530} {\bibfield
  {journal} {\bibinfo  {journal} {Phys. Rev. D}\ }\textbf {\bibinfo {volume}
  {103}},\ \bibinfo {pages} {023530} (\bibinfo {year} {2021})},\ \Eprint
  {http://arxiv.org/abs/2011.13853} {arXiv:2011.13853 [astro-ph.CO]}
  \BibitemShut {NoStop}%
\bibitem [{\citenamefont {Braglia}\ \emph {et~al.}(2021)\citenamefont
  {Braglia}, \citenamefont {Ballardini}, \citenamefont {Finelli},\ and\
  \citenamefont {Koyama}}]{Braglia:2020auw}%
  \BibitemOpen
  \bibfield  {author} {\bibinfo {author} {\bibfnamefont {M.}~\bibnamefont
  {Braglia}}, \bibinfo {author} {\bibfnamefont {M.}~\bibnamefont {Ballardini}},
  \bibinfo {author} {\bibfnamefont {F.}~\bibnamefont {Finelli}}, \ and\
  \bibinfo {author} {\bibfnamefont {K.}~\bibnamefont {Koyama}},\ }\href
  {\doibase 10.1103/PhysRevD.103.043528} {\bibfield  {journal} {\bibinfo
  {journal} {Phys. Rev. D}\ }\textbf {\bibinfo {volume} {103}},\ \bibinfo
  {pages} {043528} (\bibinfo {year} {2021})},\ \Eprint
  {http://arxiv.org/abs/2011.12934} {arXiv:2011.12934 [astro-ph.CO]}
  \BibitemShut {NoStop}%
\bibitem [{\citenamefont {Niedermann}\ and\ \citenamefont
  {Sloth}(2021)}]{Niedermann:2019olb}%
  \BibitemOpen
  \bibfield  {author} {\bibinfo {author} {\bibfnamefont {F.}~\bibnamefont
  {Niedermann}}\ and\ \bibinfo {author} {\bibfnamefont {M.~S.}\ \bibnamefont
  {Sloth}},\ }\href {\doibase 10.1103/PhysRevD.103.L041303} {\bibfield
  {journal} {\bibinfo  {journal} {Phys. Rev. D}\ }\textbf {\bibinfo {volume}
  {103}},\ \bibinfo {pages} {L041303} (\bibinfo {year} {2021})},\ \Eprint
  {http://arxiv.org/abs/1910.10739} {arXiv:1910.10739 [astro-ph.CO]}
  \BibitemShut {NoStop}%
\bibitem [{\citenamefont {Niedermann}\ and\ \citenamefont
  {Sloth}(2020)}]{Niedermann:2020dwg}%
  \BibitemOpen
  \bibfield  {author} {\bibinfo {author} {\bibfnamefont {F.}~\bibnamefont
  {Niedermann}}\ and\ \bibinfo {author} {\bibfnamefont {M.~S.}\ \bibnamefont
  {Sloth}},\ }\href {\doibase 10.1103/PhysRevD.102.063527} {\bibfield
  {journal} {\bibinfo  {journal} {Phys. Rev. D}\ }\textbf {\bibinfo {volume}
  {102}},\ \bibinfo {pages} {063527} (\bibinfo {year} {2020})},\ \Eprint
  {http://arxiv.org/abs/2006.06686} {arXiv:2006.06686 [astro-ph.CO]}
  \BibitemShut {NoStop}%
\bibitem [{\citenamefont {Cruz}\ \emph {et~al.}(2023)\citenamefont {Cruz},
  \citenamefont {Niedermann},\ and\ \citenamefont {Sloth}}]{Cruz:2023lmn}%
  \BibitemOpen
  \bibfield  {author} {\bibinfo {author} {\bibfnamefont {J.~S.}\ \bibnamefont
  {Cruz}}, \bibinfo {author} {\bibfnamefont {F.}~\bibnamefont {Niedermann}}, \
  and\ \bibinfo {author} {\bibfnamefont {M.~S.}\ \bibnamefont {Sloth}},\ }\href
  {\doibase 10.1088/1475-7516/2023/11/033} {\bibfield  {journal} {\bibinfo
  {journal} {JCAP}\ }\textbf {\bibinfo {volume} {11}},\ \bibinfo {pages} {033}
  (\bibinfo {year} {2023})},\ \Eprint {http://arxiv.org/abs/2305.08895}
  {arXiv:2305.08895 [astro-ph.CO]} \BibitemShut {NoStop}%
\bibitem [{\citenamefont {Hill}\ \emph {et~al.}(2022)\citenamefont {Hill} \emph
  {et~al.}}]{Hill:2021yec}%
  \BibitemOpen
  \bibfield  {author} {\bibinfo {author} {\bibfnamefont {J.~C.}\ \bibnamefont
  {Hill}} \emph {et~al.},\ }\href {\doibase 10.1103/PhysRevD.105.123536}
  {\bibfield  {journal} {\bibinfo  {journal} {Phys. Rev. D}\ }\textbf {\bibinfo
  {volume} {105}},\ \bibinfo {pages} {123536} (\bibinfo {year} {2022})},\
  \Eprint {http://arxiv.org/abs/2109.04451} {arXiv:2109.04451 [astro-ph.CO]}
  \BibitemShut {NoStop}%
\bibitem [{\citenamefont {Poulin}\ \emph {et~al.}(2021)\citenamefont {Poulin},
  \citenamefont {Smith},\ and\ \citenamefont {Bartlett}}]{Poulin:2021bjr}%
  \BibitemOpen
  \bibfield  {author} {\bibinfo {author} {\bibfnamefont {V.}~\bibnamefont
  {Poulin}}, \bibinfo {author} {\bibfnamefont {T.~L.}\ \bibnamefont {Smith}}, \
  and\ \bibinfo {author} {\bibfnamefont {A.}~\bibnamefont {Bartlett}},\ }\href
  {\doibase 10.1103/PhysRevD.104.123550} {\bibfield  {journal} {\bibinfo
  {journal} {Phys. Rev. D}\ }\textbf {\bibinfo {volume} {104}},\ \bibinfo
  {pages} {123550} (\bibinfo {year} {2021})},\ \Eprint
  {http://arxiv.org/abs/2109.06229} {arXiv:2109.06229 [astro-ph.CO]}
  \BibitemShut {NoStop}%
\bibitem [{\citenamefont {Smith}\ \emph
  {et~al.}(2022{\natexlab{a}})\citenamefont {Smith}, \citenamefont {Lucca},
  \citenamefont {Poulin}, \citenamefont {Abellan}, \citenamefont {Balkenhol},
  \citenamefont {Benabed}, \citenamefont {Galli},\ and\ \citenamefont
  {Murgia}}]{Smith:2022hwi}%
  \BibitemOpen
  \bibfield  {author} {\bibinfo {author} {\bibfnamefont {T.~L.}\ \bibnamefont
  {Smith}}, \bibinfo {author} {\bibfnamefont {M.}~\bibnamefont {Lucca}},
  \bibinfo {author} {\bibfnamefont {V.}~\bibnamefont {Poulin}}, \bibinfo
  {author} {\bibfnamefont {G.~F.}\ \bibnamefont {Abellan}}, \bibinfo {author}
  {\bibfnamefont {L.}~\bibnamefont {Balkenhol}}, \bibinfo {author}
  {\bibfnamefont {K.}~\bibnamefont {Benabed}}, \bibinfo {author} {\bibfnamefont
  {S.}~\bibnamefont {Galli}}, \ and\ \bibinfo {author} {\bibfnamefont
  {R.}~\bibnamefont {Murgia}},\ }\href {\doibase 10.1103/PhysRevD.106.043526}
  {\bibfield  {journal} {\bibinfo  {journal} {Phys. Rev. D}\ }\textbf {\bibinfo
  {volume} {106}},\ \bibinfo {pages} {043526} (\bibinfo {year}
  {2022}{\natexlab{a}})},\ \Eprint {http://arxiv.org/abs/2202.09379}
  {arXiv:2202.09379 [astro-ph.CO]} \BibitemShut {NoStop}%
\bibitem [{\citenamefont {Cyr-Racine}\ \emph {et~al.}(2022)\citenamefont
  {Cyr-Racine}, \citenamefont {Ge},\ and\ \citenamefont
  {Knox}}]{Cyr-Racine:2021oal}%
  \BibitemOpen
  \bibfield  {author} {\bibinfo {author} {\bibfnamefont {F.-Y.}\ \bibnamefont
  {Cyr-Racine}}, \bibinfo {author} {\bibfnamefont {F.}~\bibnamefont {Ge}}, \
  and\ \bibinfo {author} {\bibfnamefont {L.}~\bibnamefont {Knox}},\ }\href
  {\doibase 10.1103/PhysRevLett.128.201301} {\bibfield  {journal} {\bibinfo
  {journal} {Phys. Rev. Lett.}\ }\textbf {\bibinfo {volume} {128}},\ \bibinfo
  {pages} {201301} (\bibinfo {year} {2022})},\ \Eprint
  {http://arxiv.org/abs/2107.13000} {arXiv:2107.13000 [astro-ph.CO]}
  \BibitemShut {NoStop}%
\bibitem [{\citenamefont {Baryakhtar}\ \emph {et~al.}(2024)\citenamefont
  {Baryakhtar}, \citenamefont {Simon},\ and\ \citenamefont
  {Weiner}}]{Baryakhtar:2024rky}%
  \BibitemOpen
  \bibfield  {author} {\bibinfo {author} {\bibfnamefont {M.}~\bibnamefont
  {Baryakhtar}}, \bibinfo {author} {\bibfnamefont {O.}~\bibnamefont {Simon}}, \
  and\ \bibinfo {author} {\bibfnamefont {Z.~J.}\ \bibnamefont {Weiner}},\
  }\href {\doibase 10.1103/PhysRevD.110.083505} {\bibfield  {journal} {\bibinfo
   {journal} {Phys. Rev. D}\ }\textbf {\bibinfo {volume} {110}},\ \bibinfo
  {pages} {083505} (\bibinfo {year} {2024})},\ \Eprint
  {http://arxiv.org/abs/2405.10358} {arXiv:2405.10358 [astro-ph.CO]}
  \BibitemShut {NoStop}%
\bibitem [{\citenamefont {Greene}\ and\ \citenamefont
  {Cyr-Racine}(2024)}]{Greene:2024qis}%
  \BibitemOpen
  \bibfield  {author} {\bibinfo {author} {\bibfnamefont {K.}~\bibnamefont
  {Greene}}\ and\ \bibinfo {author} {\bibfnamefont {F.-Y.}\ \bibnamefont
  {Cyr-Racine}},\ }\href {\doibase 10.1103/PhysRevD.110.043524} {\bibfield
  {journal} {\bibinfo  {journal} {Phys. Rev. D}\ }\textbf {\bibinfo {volume}
  {110}},\ \bibinfo {pages} {043524} (\bibinfo {year} {2024})},\ \Eprint
  {http://arxiv.org/abs/2403.05619} {arXiv:2403.05619 [astro-ph.CO]}
  \BibitemShut {NoStop}%
\bibitem [{\citenamefont {Benevento}\ \emph {et~al.}(2020)\citenamefont
  {Benevento}, \citenamefont {Hu},\ and\ \citenamefont
  {Raveri}}]{Benevento:2020fev}%
  \BibitemOpen
  \bibfield  {author} {\bibinfo {author} {\bibfnamefont {G.}~\bibnamefont
  {Benevento}}, \bibinfo {author} {\bibfnamefont {W.}~\bibnamefont {Hu}}, \
  and\ \bibinfo {author} {\bibfnamefont {M.}~\bibnamefont {Raveri}},\ }\href
  {\doibase 10.1103/PhysRevD.101.103517} {\bibfield  {journal} {\bibinfo
  {journal} {Phys. Rev. D}\ }\textbf {\bibinfo {volume} {101}},\ \bibinfo
  {pages} {103517} (\bibinfo {year} {2020})},\ \Eprint
  {http://arxiv.org/abs/2002.11707} {arXiv:2002.11707 [astro-ph.CO]}
  \BibitemShut {NoStop}%
\bibitem [{\citenamefont {Camarena}\ and\ \citenamefont
  {Marra}(2021)}]{Camarena:2021jlr}%
  \BibitemOpen
  \bibfield  {author} {\bibinfo {author} {\bibfnamefont {D.}~\bibnamefont
  {Camarena}}\ and\ \bibinfo {author} {\bibfnamefont {V.}~\bibnamefont
  {Marra}},\ }\href {\doibase 10.1093/mnras/stab1200} {\bibfield  {journal}
  {\bibinfo  {journal} {Mon. Not. Roy. Astron. Soc.}\ }\textbf {\bibinfo
  {volume} {504}},\ \bibinfo {pages} {5164} (\bibinfo {year} {2021})},\ \Eprint
  {http://arxiv.org/abs/2101.08641} {arXiv:2101.08641 [astro-ph.CO]}
  \BibitemShut {NoStop}%
\bibitem [{\citenamefont {Alam}\ \emph {et~al.}(2021)\citenamefont {Alam} \emph
  {et~al.}}]{eBOSS:2020yzd}%
  \BibitemOpen
  \bibfield  {author} {\bibinfo {author} {\bibfnamefont {S.}~\bibnamefont
  {Alam}} \emph {et~al.} (\bibinfo {collaboration} {eBOSS}),\ }\href {\doibase
  10.1103/PhysRevD.103.083533} {\bibfield  {journal} {\bibinfo  {journal}
  {Phys. Rev. D}\ }\textbf {\bibinfo {volume} {103}},\ \bibinfo {pages}
  {083533} (\bibinfo {year} {2021})},\ \Eprint
  {http://arxiv.org/abs/2007.08991} {arXiv:2007.08991 [astro-ph.CO]}
  \BibitemShut {NoStop}%
\bibitem [{\citenamefont {Abdul~Karim}\ \emph {et~al.}(2025)\citenamefont
  {Abdul~Karim} \emph {et~al.}}]{DESI:2025zgx}%
  \BibitemOpen
  \bibfield  {author} {\bibinfo {author} {\bibfnamefont {M.}~\bibnamefont
  {Abdul~Karim}} \emph {et~al.} (\bibinfo {collaboration} {DESI}),\ }\href@noop
  {} {\  (\bibinfo {year} {2025})},\ \Eprint {http://arxiv.org/abs/2503.14738}
  {arXiv:2503.14738 [astro-ph.CO]} \BibitemShut {NoStop}%
\bibitem [{\citenamefont {Chaussidon}\ \emph {et~al.}(2025)\citenamefont
  {Chaussidon} \emph {et~al.}}]{Chaussidon:2025tww}%
  \BibitemOpen
  \bibfield  {author} {\bibinfo {author} {\bibfnamefont {E.}~\bibnamefont
  {Chaussidon}} \emph {et~al.},\ }\href@noop {} {\  (\bibinfo {year} {2025})},\
  \Eprint {http://arxiv.org/abs/2503.24343} {arXiv:2503.24343 [astro-ph.CO]}
  \BibitemShut {NoStop}%
\bibitem [{\citenamefont {Abbott}\ \emph {et~al.}(2024)\citenamefont {Abbott}
  \emph {et~al.}}]{DES:2024jxu}%
  \BibitemOpen
  \bibfield  {author} {\bibinfo {author} {\bibfnamefont {T.~M.~C.}\
  \bibnamefont {Abbott}} \emph {et~al.} (\bibinfo {collaboration} {DES}),\
  }\href {\doibase 10.3847/2041-8213/ad6f9f} {\bibfield  {journal} {\bibinfo
  {journal} {Astrophys. J. Lett.}\ }\textbf {\bibinfo {volume} {973}},\
  \bibinfo {pages} {L14} (\bibinfo {year} {2024})},\ \Eprint
  {http://arxiv.org/abs/2401.02929} {arXiv:2401.02929 [astro-ph.CO]}
  \BibitemShut {NoStop}%
\bibitem [{\citenamefont {Rubin}\ \emph {et~al.}(2023)\citenamefont {Rubin}
  \emph {et~al.}}]{Rubin:2023ovl}%
  \BibitemOpen
  \bibfield  {author} {\bibinfo {author} {\bibfnamefont {D.}~\bibnamefont
  {Rubin}} \emph {et~al.},\ }\href@noop {} {\  (\bibinfo {year} {2023})},\
  \Eprint {http://arxiv.org/abs/2311.12098} {arXiv:2311.12098 [astro-ph.CO]}
  \BibitemShut {NoStop}%
\bibitem [{\citenamefont {Trendafilova}\ \emph {et~al.}(2025)\citenamefont
  {Trendafilova}, \citenamefont {Khalife},\ and\ \citenamefont
  {Galli}}]{Trendafilova:2025dce}%
  \BibitemOpen
  \bibfield  {author} {\bibinfo {author} {\bibfnamefont {C.}~\bibnamefont
  {Trendafilova}}, \bibinfo {author} {\bibfnamefont {A.~R.}\ \bibnamefont
  {Khalife}}, \ and\ \bibinfo {author} {\bibfnamefont {S.}~\bibnamefont
  {Galli}},\ }\href@noop {} {\  (\bibinfo {year} {2025})},\ \Eprint
  {http://arxiv.org/abs/2502.19383} {arXiv:2502.19383 [astro-ph.CO]}
  \BibitemShut {NoStop}%
\bibitem [{\citenamefont {Audren}\ \emph {et~al.}(2013)\citenamefont {Audren},
  \citenamefont {Lesgourgues}, \citenamefont {Benabed},\ and\ \citenamefont
  {Prunet}}]{Audren:2012wb}%
  \BibitemOpen
  \bibfield  {author} {\bibinfo {author} {\bibfnamefont {B.}~\bibnamefont
  {Audren}}, \bibinfo {author} {\bibfnamefont {J.}~\bibnamefont {Lesgourgues}},
  \bibinfo {author} {\bibfnamefont {K.}~\bibnamefont {Benabed}}, \ and\
  \bibinfo {author} {\bibfnamefont {S.}~\bibnamefont {Prunet}},\ }\href
  {\doibase 10.1088/1475-7516/2013/02/001} {\bibfield  {journal} {\bibinfo
  {journal} {JCAP}\ }\textbf {\bibinfo {volume} {1302}},\ \bibinfo {pages}
  {001} (\bibinfo {year} {2013})},\ \Eprint {http://arxiv.org/abs/1210.7183}
  {arXiv:1210.7183 [astro-ph.CO]} \BibitemShut {NoStop}%
\bibitem [{\citenamefont {Brinckmann}\ and\ \citenamefont
  {Lesgourgues}(2018)}]{Brinckmann:2018cvx}%
  \BibitemOpen
  \bibfield  {author} {\bibinfo {author} {\bibfnamefont {T.}~\bibnamefont
  {Brinckmann}}\ and\ \bibinfo {author} {\bibfnamefont {J.}~\bibnamefont
  {Lesgourgues}},\ }\href@noop {} {\  (\bibinfo {year} {2018})},\ \Eprint
  {http://arxiv.org/abs/1804.07261} {arXiv:1804.07261 [astro-ph.CO]}
  \BibitemShut {NoStop}%
\bibitem [{\citenamefont {Aghanim}\ \emph
  {et~al.}(2020{\natexlab{b}})\citenamefont {Aghanim} \emph
  {et~al.}}]{Planck:2019nip}%
  \BibitemOpen
  \bibfield  {author} {\bibinfo {author} {\bibfnamefont {N.}~\bibnamefont
  {Aghanim}} \emph {et~al.} (\bibinfo {collaboration} {Planck}),\ }\href
  {\doibase 10.1051/0004-6361/201936386} {\bibfield  {journal} {\bibinfo
  {journal} {Astron. Astrophys.}\ }\textbf {\bibinfo {volume} {641}},\ \bibinfo
  {pages} {A5} (\bibinfo {year} {2020}{\natexlab{b}})},\ \Eprint
  {http://arxiv.org/abs/1907.12875} {arXiv:1907.12875 [astro-ph.CO]}
  \BibitemShut {NoStop}%
\bibitem [{\citenamefont {Akrami}\ \emph {et~al.}(2020)\citenamefont {Akrami}
  \emph {et~al.}}]{Planck:2020olo}%
  \BibitemOpen
  \bibfield  {author} {\bibinfo {author} {\bibfnamefont {Y.}~\bibnamefont
  {Akrami}} \emph {et~al.} (\bibinfo {collaboration} {Planck}),\ }\href
  {\doibase 10.1051/0004-6361/202038073} {\bibfield  {journal} {\bibinfo
  {journal} {Astron. Astrophys.}\ }\textbf {\bibinfo {volume} {643}},\ \bibinfo
  {pages} {A42} (\bibinfo {year} {2020})},\ \Eprint
  {http://arxiv.org/abs/2007.04997} {arXiv:2007.04997 [astro-ph.CO]}
  \BibitemShut {NoStop}%
\bibitem [{\citenamefont {Rosenberg}\ \emph {et~al.}(2022)\citenamefont
  {Rosenberg}, \citenamefont {Gratton},\ and\ \citenamefont
  {Efstathiou}}]{Rosenberg:2022sdy}%
  \BibitemOpen
  \bibfield  {author} {\bibinfo {author} {\bibfnamefont {E.}~\bibnamefont
  {Rosenberg}}, \bibinfo {author} {\bibfnamefont {S.}~\bibnamefont {Gratton}},
  \ and\ \bibinfo {author} {\bibfnamefont {G.}~\bibnamefont {Efstathiou}},\
  }\href {\doibase 10.1093/mnras/stac2744} {\bibfield  {journal} {\bibinfo
  {journal} {Mon. Not. Roy. Astron. Soc.}\ }\textbf {\bibinfo {volume} {517}},\
  \bibinfo {pages} {4620} (\bibinfo {year} {2022})},\ \Eprint
  {http://arxiv.org/abs/2205.10869} {arXiv:2205.10869 [astro-ph.CO]}
  \BibitemShut {NoStop}%
\bibitem [{\citenamefont {Alam}\ \emph {et~al.}(2017)\citenamefont {Alam} \emph
  {et~al.}}]{BOSS:2016wmc}%
  \BibitemOpen
  \bibfield  {author} {\bibinfo {author} {\bibfnamefont {S.}~\bibnamefont
  {Alam}} \emph {et~al.} (\bibinfo {collaboration} {BOSS}),\ }\href {\doibase
  10.1093/mnras/stx721} {\bibfield  {journal} {\bibinfo  {journal} {Mon. Not.
  Roy. Astron. Soc.}\ }\textbf {\bibinfo {volume} {470}},\ \bibinfo {pages}
  {2617} (\bibinfo {year} {2017})},\ \Eprint {http://arxiv.org/abs/1607.03155}
  {arXiv:1607.03155 [astro-ph.CO]} \BibitemShut {NoStop}%
\bibitem [{\citenamefont {Ross}\ \emph {et~al.}(2015)\citenamefont {Ross},
  \citenamefont {Samushia}, \citenamefont {Howlett}, \citenamefont {Percival},
  \citenamefont {Burden},\ and\ \citenamefont {Manera}}]{Ross:2014qpa}%
  \BibitemOpen
  \bibfield  {author} {\bibinfo {author} {\bibfnamefont {A.~J.}\ \bibnamefont
  {Ross}}, \bibinfo {author} {\bibfnamefont {L.}~\bibnamefont {Samushia}},
  \bibinfo {author} {\bibfnamefont {C.}~\bibnamefont {Howlett}}, \bibinfo
  {author} {\bibfnamefont {W.~J.}\ \bibnamefont {Percival}}, \bibinfo {author}
  {\bibfnamefont {A.}~\bibnamefont {Burden}}, \ and\ \bibinfo {author}
  {\bibfnamefont {M.}~\bibnamefont {Manera}},\ }\href {\doibase
  10.1093/mnras/stv154} {\bibfield  {journal} {\bibinfo  {journal} {Mon. Not.
  Roy. Astron. Soc.}\ }\textbf {\bibinfo {volume} {449}},\ \bibinfo {pages}
  {835} (\bibinfo {year} {2015})},\ \Eprint {http://arxiv.org/abs/1409.3242}
  {arXiv:1409.3242 [astro-ph.CO]} \BibitemShut {NoStop}%
\bibitem [{\citenamefont {{Beutler}}\ \emph {et~al.}(2011)\citenamefont
  {{Beutler}}, \citenamefont {{Blake}}, \citenamefont {{Colless}},
  \citenamefont {{Jones}}, \citenamefont {{Staveley-Smith}}, \citenamefont
  {{Campbell}}, \citenamefont {{Parker}}, \citenamefont {{Saunders}},\ and\
  \citenamefont {{Watson}}}]{2011MNRAS.416.3017B}%
  \BibitemOpen
  \bibfield  {author} {\bibinfo {author} {\bibfnamefont {F.}~\bibnamefont
  {{Beutler}}}, \bibinfo {author} {\bibfnamefont {C.}~\bibnamefont {{Blake}}},
  \bibinfo {author} {\bibfnamefont {M.}~\bibnamefont {{Colless}}}, \bibinfo
  {author} {\bibfnamefont {D.~H.}\ \bibnamefont {{Jones}}}, \bibinfo {author}
  {\bibfnamefont {L.}~\bibnamefont {{Staveley-Smith}}}, \bibinfo {author}
  {\bibfnamefont {L.}~\bibnamefont {{Campbell}}}, \bibinfo {author}
  {\bibfnamefont {Q.}~\bibnamefont {{Parker}}}, \bibinfo {author}
  {\bibfnamefont {W.}~\bibnamefont {{Saunders}}}, \ and\ \bibinfo {author}
  {\bibfnamefont {F.}~\bibnamefont {{Watson}}},\ }\href {\doibase
  10.1111/j.1365-2966.2011.19250.x} {\bibfield  {journal} {\bibinfo  {journal}
  {\mnras}\ }\textbf {\bibinfo {volume} {416}},\ \bibinfo {pages} {3017}
  (\bibinfo {year} {2011})},\ \Eprint {http://arxiv.org/abs/1106.3366}
  {arXiv:1106.3366 [astro-ph.CO]} \BibitemShut {NoStop}%
\bibitem [{\citenamefont {Follin}\ \emph {et~al.}(2015)\citenamefont {Follin},
  \citenamefont {Knox}, \citenamefont {Millea},\ and\ \citenamefont
  {Pan}}]{Follin:2015hya}%
  \BibitemOpen
  \bibfield  {author} {\bibinfo {author} {\bibfnamefont {B.}~\bibnamefont
  {Follin}}, \bibinfo {author} {\bibfnamefont {L.}~\bibnamefont {Knox}},
  \bibinfo {author} {\bibfnamefont {M.}~\bibnamefont {Millea}}, \ and\ \bibinfo
  {author} {\bibfnamefont {Z.}~\bibnamefont {Pan}},\ }\href {\doibase
  10.1103/PhysRevLett.115.091301} {\bibfield  {journal} {\bibinfo  {journal}
  {Phys. Rev. Lett.}\ }\textbf {\bibinfo {volume} {115}},\ \bibinfo {pages}
  {091301} (\bibinfo {year} {2015})},\ \Eprint
  {http://arxiv.org/abs/1503.07863} {arXiv:1503.07863 [astro-ph.CO]}
  \BibitemShut {NoStop}%
\bibitem [{\citenamefont {Baumann}\ \emph {et~al.}(2019)\citenamefont
  {Baumann}, \citenamefont {Beutler}, \citenamefont {Flauger}, \citenamefont
  {Green}, \citenamefont {Slosar}, \citenamefont {Vargas-Maga\~na},
  \citenamefont {Wallisch},\ and\ \citenamefont {Y\`eche}}]{Baumann:2019keh}%
  \BibitemOpen
  \bibfield  {author} {\bibinfo {author} {\bibfnamefont {D.~D.}\ \bibnamefont
  {Baumann}}, \bibinfo {author} {\bibfnamefont {F.}~\bibnamefont {Beutler}},
  \bibinfo {author} {\bibfnamefont {R.}~\bibnamefont {Flauger}}, \bibinfo
  {author} {\bibfnamefont {D.~R.}\ \bibnamefont {Green}}, \bibinfo {author}
  {\bibfnamefont {A.}~\bibnamefont {Slosar}}, \bibinfo {author} {\bibfnamefont
  {M.}~\bibnamefont {Vargas-Maga\~na}}, \bibinfo {author} {\bibfnamefont
  {B.}~\bibnamefont {Wallisch}}, \ and\ \bibinfo {author} {\bibfnamefont
  {C.}~\bibnamefont {Y\`eche}},\ }\href {\doibase 10.1038/s41567-019-0435-6}
  {\bibfield  {journal} {\bibinfo  {journal} {Nature Phys.}\ }\textbf {\bibinfo
  {volume} {15}},\ \bibinfo {pages} {465} (\bibinfo {year} {2019})},\ \Eprint
  {http://arxiv.org/abs/1803.10741} {arXiv:1803.10741 [astro-ph.CO]}
  \BibitemShut {NoStop}%
\bibitem [{\citenamefont {Saravanan}\ \emph {et~al.}(2025)\citenamefont
  {Saravanan}, \citenamefont {Brinckmann}, \citenamefont {Loverde},\ and\
  \citenamefont {Weiner}}]{Saravanan:2025cyi}%
  \BibitemOpen
  \bibfield  {author} {\bibinfo {author} {\bibfnamefont {M.~M.}\ \bibnamefont
  {Saravanan}}, \bibinfo {author} {\bibfnamefont {T.}~\bibnamefont
  {Brinckmann}}, \bibinfo {author} {\bibfnamefont {M.}~\bibnamefont {Loverde}},
  \ and\ \bibinfo {author} {\bibfnamefont {Z.~J.}\ \bibnamefont {Weiner}},\
  }\href@noop {} {\  (\bibinfo {year} {2025})},\ \Eprint
  {http://arxiv.org/abs/2503.04671} {arXiv:2503.04671 [astro-ph.CO]}
  \BibitemShut {NoStop}%
\bibitem [{\citenamefont {Montefalcone}\ \emph {et~al.}(2025)\citenamefont
  {Montefalcone}, \citenamefont {Wallisch},\ and\ \citenamefont
  {Freese}}]{Montefalcone:2025unv}%
  \BibitemOpen
  \bibfield  {author} {\bibinfo {author} {\bibfnamefont {G.}~\bibnamefont
  {Montefalcone}}, \bibinfo {author} {\bibfnamefont {B.}~\bibnamefont
  {Wallisch}}, \ and\ \bibinfo {author} {\bibfnamefont {K.}~\bibnamefont
  {Freese}},\ }\href@noop {} {\  (\bibinfo {year} {2025})},\ \Eprint
  {http://arxiv.org/abs/2501.13788} {arXiv:2501.13788 [astro-ph.CO]}
  \BibitemShut {NoStop}%
\bibitem [{\citenamefont {Blinov}\ and\ \citenamefont
  {Marques-Tavares}(2020)}]{Blinov:2020hmc}%
  \BibitemOpen
  \bibfield  {author} {\bibinfo {author} {\bibfnamefont {N.}~\bibnamefont
  {Blinov}}\ and\ \bibinfo {author} {\bibfnamefont {G.}~\bibnamefont
  {Marques-Tavares}},\ }\href {\doibase 10.1088/1475-7516/2020/09/029}
  {\bibfield  {journal} {\bibinfo  {journal} {JCAP}\ }\textbf {\bibinfo
  {volume} {09}},\ \bibinfo {pages} {029} (\bibinfo {year} {2020})},\ \Eprint
  {http://arxiv.org/abs/2003.08387} {arXiv:2003.08387 [astro-ph.CO]}
  \BibitemShut {NoStop}%
\bibitem [{\citenamefont {Brust}\ \emph {et~al.}(2017)\citenamefont {Brust},
  \citenamefont {Cui},\ and\ \citenamefont {Sigurdson}}]{Brust:2017nmv}%
  \BibitemOpen
  \bibfield  {author} {\bibinfo {author} {\bibfnamefont {C.}~\bibnamefont
  {Brust}}, \bibinfo {author} {\bibfnamefont {Y.}~\bibnamefont {Cui}}, \ and\
  \bibinfo {author} {\bibfnamefont {K.}~\bibnamefont {Sigurdson}},\ }\href
  {\doibase 10.1088/1475-7516/2017/08/020} {\bibfield  {journal} {\bibinfo
  {journal} {JCAP}\ }\textbf {\bibinfo {volume} {08}},\ \bibinfo {pages} {020}
  (\bibinfo {year} {2017})},\ \Eprint {http://arxiv.org/abs/1703.10732}
  {arXiv:1703.10732 [astro-ph.CO]} \BibitemShut {NoStop}%
\bibitem [{\citenamefont {Becker}\ \emph {et~al.}(2021)\citenamefont {Becker},
  \citenamefont {Hooper}, \citenamefont {Kahlhoefer}, \citenamefont
  {Lesgourgues},\ and\ \citenamefont {Sch\"oneberg}}]{Becker:2020hzj}%
  \BibitemOpen
  \bibfield  {author} {\bibinfo {author} {\bibfnamefont {N.}~\bibnamefont
  {Becker}}, \bibinfo {author} {\bibfnamefont {D.~C.}\ \bibnamefont {Hooper}},
  \bibinfo {author} {\bibfnamefont {F.}~\bibnamefont {Kahlhoefer}}, \bibinfo
  {author} {\bibfnamefont {J.}~\bibnamefont {Lesgourgues}}, \ and\ \bibinfo
  {author} {\bibfnamefont {N.}~\bibnamefont {Sch\"oneberg}},\ }\href {\doibase
  10.1088/1475-7516/2021/02/019} {\bibfield  {journal} {\bibinfo  {journal}
  {JCAP}\ }\textbf {\bibinfo {volume} {02}},\ \bibinfo {pages} {019} (\bibinfo
  {year} {2021})},\ \Eprint {http://arxiv.org/abs/2010.04074} {arXiv:2010.04074
  [astro-ph.CO]} \BibitemShut {NoStop}%
\bibitem [{\citenamefont {Archidiacono}\ \emph {et~al.}(2019)\citenamefont
  {Archidiacono}, \citenamefont {Hooper}, \citenamefont {Murgia}, \citenamefont
  {Bohr}, \citenamefont {Lesgourgues},\ and\ \citenamefont
  {Viel}}]{Archidiacono:2019wdp}%
  \BibitemOpen
  \bibfield  {author} {\bibinfo {author} {\bibfnamefont {M.}~\bibnamefont
  {Archidiacono}}, \bibinfo {author} {\bibfnamefont {D.~C.}\ \bibnamefont
  {Hooper}}, \bibinfo {author} {\bibfnamefont {R.}~\bibnamefont {Murgia}},
  \bibinfo {author} {\bibfnamefont {S.}~\bibnamefont {Bohr}}, \bibinfo {author}
  {\bibfnamefont {J.}~\bibnamefont {Lesgourgues}}, \ and\ \bibinfo {author}
  {\bibfnamefont {M.}~\bibnamefont {Viel}},\ }\href {\doibase
  10.1088/1475-7516/2019/10/055} {\bibfield  {journal} {\bibinfo  {journal}
  {JCAP}\ }\textbf {\bibinfo {volume} {10}},\ \bibinfo {pages} {055} (\bibinfo
  {year} {2019})},\ \Eprint {http://arxiv.org/abs/1907.01496} {arXiv:1907.01496
  [astro-ph.CO]} \BibitemShut {NoStop}%
\bibitem [{\citenamefont {Aloni}\ \emph {et~al.}(2023)\citenamefont {Aloni},
  \citenamefont {Joseph}, \citenamefont {Schmaltz},\ and\ \citenamefont
  {Weiner}}]{Aloni:2023tff}%
  \BibitemOpen
  \bibfield  {author} {\bibinfo {author} {\bibfnamefont {D.}~\bibnamefont
  {Aloni}}, \bibinfo {author} {\bibfnamefont {M.}~\bibnamefont {Joseph}},
  \bibinfo {author} {\bibfnamefont {M.}~\bibnamefont {Schmaltz}}, \ and\
  \bibinfo {author} {\bibfnamefont {N.}~\bibnamefont {Weiner}},\ }\href
  {\doibase 10.1103/PhysRevLett.131.221001} {\bibfield  {journal} {\bibinfo
  {journal} {Phys. Rev. Lett.}\ }\textbf {\bibinfo {volume} {131}},\ \bibinfo
  {pages} {221001} (\bibinfo {year} {2023})},\ \Eprint
  {http://arxiv.org/abs/2301.10792} {arXiv:2301.10792 [astro-ph.CO]}
  \BibitemShut {NoStop}%
\bibitem [{\citenamefont {Joseph}\ \emph {et~al.}(2023)\citenamefont {Joseph},
  \citenamefont {Aloni}, \citenamefont {Schmaltz}, \citenamefont {Sivarajan},\
  and\ \citenamefont {Weiner}}]{Joseph:2022jsf}%
  \BibitemOpen
  \bibfield  {author} {\bibinfo {author} {\bibfnamefont {M.}~\bibnamefont
  {Joseph}}, \bibinfo {author} {\bibfnamefont {D.}~\bibnamefont {Aloni}},
  \bibinfo {author} {\bibfnamefont {M.}~\bibnamefont {Schmaltz}}, \bibinfo
  {author} {\bibfnamefont {E.~N.}\ \bibnamefont {Sivarajan}}, \ and\ \bibinfo
  {author} {\bibfnamefont {N.}~\bibnamefont {Weiner}},\ }\href {\doibase
  10.1103/PhysRevD.108.023520} {\bibfield  {journal} {\bibinfo  {journal}
  {Phys. Rev. D}\ }\textbf {\bibinfo {volume} {108}},\ \bibinfo {pages}
  {023520} (\bibinfo {year} {2023})},\ \Eprint
  {http://arxiv.org/abs/2207.03500} {arXiv:2207.03500 [astro-ph.CO]}
  \BibitemShut {NoStop}%
\bibitem [{\citenamefont {Poulin}\ \emph {et~al.}(2023)\citenamefont {Poulin},
  \citenamefont {Smith},\ and\ \citenamefont {Karwal}}]{Poulin:2023lkg}%
  \BibitemOpen
  \bibfield  {author} {\bibinfo {author} {\bibfnamefont {V.}~\bibnamefont
  {Poulin}}, \bibinfo {author} {\bibfnamefont {T.~L.}\ \bibnamefont {Smith}}, \
  and\ \bibinfo {author} {\bibfnamefont {T.}~\bibnamefont {Karwal}},\ }\href
  {\doibase 10.1016/j.dark.2023.101348} {\bibfield  {journal} {\bibinfo
  {journal} {Phys. Dark Univ.}\ }\textbf {\bibinfo {volume} {42}},\ \bibinfo
  {pages} {101348} (\bibinfo {year} {2023})},\ \Eprint
  {http://arxiv.org/abs/2302.09032} {arXiv:2302.09032 [astro-ph.CO]}
  \BibitemShut {NoStop}%
\bibitem [{\citenamefont {Simon}\ \emph {et~al.}(2023)\citenamefont {Simon},
  \citenamefont {Zhang}, \citenamefont {Poulin},\ and\ \citenamefont
  {Smith}}]{Simon:2022adh}%
  \BibitemOpen
  \bibfield  {author} {\bibinfo {author} {\bibfnamefont {T.}~\bibnamefont
  {Simon}}, \bibinfo {author} {\bibfnamefont {P.}~\bibnamefont {Zhang}},
  \bibinfo {author} {\bibfnamefont {V.}~\bibnamefont {Poulin}}, \ and\ \bibinfo
  {author} {\bibfnamefont {T.~L.}\ \bibnamefont {Smith}},\ }\href {\doibase
  10.1103/PhysRevD.107.063505} {\bibfield  {journal} {\bibinfo  {journal}
  {Phys. Rev. D}\ }\textbf {\bibinfo {volume} {107}},\ \bibinfo {pages}
  {063505} (\bibinfo {year} {2023})},\ \Eprint
  {http://arxiv.org/abs/2208.05930} {arXiv:2208.05930 [astro-ph.CO]}
  \BibitemShut {NoStop}%
\bibitem [{\citenamefont {Smith}\ and\ \citenamefont
  {Poulin}(2024)}]{Smith:2023oop}%
  \BibitemOpen
  \bibfield  {author} {\bibinfo {author} {\bibfnamefont {T.~L.}\ \bibnamefont
  {Smith}}\ and\ \bibinfo {author} {\bibfnamefont {V.}~\bibnamefont {Poulin}},\
  }\href {\doibase 10.1103/PhysRevD.109.103506} {\bibfield  {journal} {\bibinfo
   {journal} {Phys. Rev. D}\ }\textbf {\bibinfo {volume} {109}},\ \bibinfo
  {pages} {103506} (\bibinfo {year} {2024})},\ \Eprint
  {http://arxiv.org/abs/2309.03265} {arXiv:2309.03265 [astro-ph.CO]}
  \BibitemShut {NoStop}%
\bibitem [{\citenamefont {Efstathiou}\ \emph {et~al.}(2024)\citenamefont
  {Efstathiou}, \citenamefont {Rosenberg},\ and\ \citenamefont
  {Poulin}}]{Efstathiou:2023fbn}%
  \BibitemOpen
  \bibfield  {author} {\bibinfo {author} {\bibfnamefont {G.}~\bibnamefont
  {Efstathiou}}, \bibinfo {author} {\bibfnamefont {E.}~\bibnamefont
  {Rosenberg}}, \ and\ \bibinfo {author} {\bibfnamefont {V.}~\bibnamefont
  {Poulin}},\ }\href {\doibase 10.1103/PhysRevLett.132.221002} {\bibfield
  {journal} {\bibinfo  {journal} {Phys. Rev. Lett.}\ }\textbf {\bibinfo
  {volume} {132}},\ \bibinfo {pages} {221002} (\bibinfo {year} {2024})},\
  \Eprint {http://arxiv.org/abs/2311.00524} {arXiv:2311.00524 [astro-ph.CO]}
  \BibitemShut {NoStop}%
\bibitem [{\citenamefont {Smith}\ \emph {et~al.}(2020)\citenamefont {Smith},
  \citenamefont {Poulin},\ and\ \citenamefont {Amin}}]{Smith:2019ihp}%
  \BibitemOpen
  \bibfield  {author} {\bibinfo {author} {\bibfnamefont {T.~L.}\ \bibnamefont
  {Smith}}, \bibinfo {author} {\bibfnamefont {V.}~\bibnamefont {Poulin}}, \
  and\ \bibinfo {author} {\bibfnamefont {M.~A.}\ \bibnamefont {Amin}},\ }\href
  {\doibase 10.1103/PhysRevD.101.063523} {\bibfield  {journal} {\bibinfo
  {journal} {Phys. Rev. D}\ }\textbf {\bibinfo {volume} {101}},\ \bibinfo
  {pages} {063523} (\bibinfo {year} {2020})},\ \Eprint
  {http://arxiv.org/abs/1908.06995} {arXiv:1908.06995 [astro-ph.CO]}
  \BibitemShut {NoStop}%
\bibitem [{\citenamefont {Turner}(1983)}]{Turner:1983he}%
  \BibitemOpen
  \bibfield  {author} {\bibinfo {author} {\bibfnamefont {M.~S.}\ \bibnamefont
  {Turner}},\ }\href {\doibase 10.1103/PhysRevD.28.1243} {\bibfield  {journal}
  {\bibinfo  {journal} {Phys. Rev. D}\ }\textbf {\bibinfo {volume} {28}},\
  \bibinfo {pages} {1243} (\bibinfo {year} {1983})}\BibitemShut {NoStop}%
\bibitem [{\citenamefont {Simon}(2024)}]{Simon:2023hlp}%
  \BibitemOpen
  \bibfield  {author} {\bibinfo {author} {\bibfnamefont {T.}~\bibnamefont
  {Simon}},\ }\href {\doibase 10.1103/PhysRevD.110.023528} {\bibfield
  {journal} {\bibinfo  {journal} {Phys. Rev. D}\ }\textbf {\bibinfo {volume}
  {110}},\ \bibinfo {pages} {023528} (\bibinfo {year} {2024})},\ \Eprint
  {http://arxiv.org/abs/2310.16800} {arXiv:2310.16800 [astro-ph.CO]}
  \BibitemShut {NoStop}%
\bibitem [{\citenamefont {Lin}\ \emph {et~al.}(2019)\citenamefont {Lin},
  \citenamefont {Benevento}, \citenamefont {Hu},\ and\ \citenamefont
  {Raveri}}]{Lin:2019qug}%
  \BibitemOpen
  \bibfield  {author} {\bibinfo {author} {\bibfnamefont {M.-X.}\ \bibnamefont
  {Lin}}, \bibinfo {author} {\bibfnamefont {G.}~\bibnamefont {Benevento}},
  \bibinfo {author} {\bibfnamefont {W.}~\bibnamefont {Hu}}, \ and\ \bibinfo
  {author} {\bibfnamefont {M.}~\bibnamefont {Raveri}},\ }\href {\doibase
  10.1103/PhysRevD.100.063542} {\bibfield  {journal} {\bibinfo  {journal}
  {Phys. Rev. D}\ }\textbf {\bibinfo {volume} {100}},\ \bibinfo {pages}
  {063542} (\bibinfo {year} {2019})},\ \Eprint
  {http://arxiv.org/abs/1905.12618} {arXiv:1905.12618 [astro-ph.CO]}
  \BibitemShut {NoStop}%
\bibitem [{\citenamefont {Chatrchyan}\ \emph {et~al.}(2025)\citenamefont
  {Chatrchyan}, \citenamefont {Niedermann}, \citenamefont {Poulin},\ and\
  \citenamefont {Sloth}}]{Chatrchyan:2024xjj}%
  \BibitemOpen
  \bibfield  {author} {\bibinfo {author} {\bibfnamefont {A.}~\bibnamefont
  {Chatrchyan}}, \bibinfo {author} {\bibfnamefont {F.}~\bibnamefont
  {Niedermann}}, \bibinfo {author} {\bibfnamefont {V.}~\bibnamefont {Poulin}},
  \ and\ \bibinfo {author} {\bibfnamefont {M.~S.}\ \bibnamefont {Sloth}},\
  }\href {\doibase 10.1103/PhysRevD.111.043536} {\bibfield  {journal} {\bibinfo
   {journal} {Phys. Rev. D}\ }\textbf {\bibinfo {volume} {111}},\ \bibinfo
  {pages} {043536} (\bibinfo {year} {2025})},\ \Eprint
  {http://arxiv.org/abs/2408.14537} {arXiv:2408.14537 [astro-ph.CO]}
  \BibitemShut {NoStop}%
\bibitem [{\citenamefont {Sekiguchi}\ and\ \citenamefont
  {Takahashi}(2021)}]{Sekiguchi:2020teg}%
  \BibitemOpen
  \bibfield  {author} {\bibinfo {author} {\bibfnamefont {T.}~\bibnamefont
  {Sekiguchi}}\ and\ \bibinfo {author} {\bibfnamefont {T.}~\bibnamefont
  {Takahashi}},\ }\href {\doibase 10.1103/PhysRevD.103.083507} {\bibfield
  {journal} {\bibinfo  {journal} {Phys. Rev. D}\ }\textbf {\bibinfo {volume}
  {103}},\ \bibinfo {pages} {083507} (\bibinfo {year} {2021})},\ \Eprint
  {http://arxiv.org/abs/2007.03381} {arXiv:2007.03381 [astro-ph.CO]}
  \BibitemShut {NoStop}%
\bibitem [{\citenamefont {Baryakhtar}\ \emph {et~al.}(2025)\citenamefont
  {Baryakhtar}, \citenamefont {Simon},\ and\ \citenamefont
  {Weiner}}]{Baryakhtar:2025uxs}%
  \BibitemOpen
  \bibfield  {author} {\bibinfo {author} {\bibfnamefont {M.}~\bibnamefont
  {Baryakhtar}}, \bibinfo {author} {\bibfnamefont {O.}~\bibnamefont {Simon}}, \
  and\ \bibinfo {author} {\bibfnamefont {Z.~J.}\ \bibnamefont {Weiner}},\
  }\href@noop {} {\  (\bibinfo {year} {2025})},\ \Eprint
  {http://arxiv.org/abs/2502.04432} {arXiv:2502.04432 [hep-ph]} \BibitemShut
  {NoStop}%
\bibitem [{\citenamefont {Hart}\ and\ \citenamefont
  {Chluba}(2018)}]{Hart:2017ndk}%
  \BibitemOpen
  \bibfield  {author} {\bibinfo {author} {\bibfnamefont {L.}~\bibnamefont
  {Hart}}\ and\ \bibinfo {author} {\bibfnamefont {J.}~\bibnamefont {Chluba}},\
  }\href {\doibase 10.1093/mnras/stx2783} {\bibfield  {journal} {\bibinfo
  {journal} {Mon. Not. Roy. Astron. Soc.}\ }\textbf {\bibinfo {volume} {474}},\
  \bibinfo {pages} {1850} (\bibinfo {year} {2018})},\ \Eprint
  {http://arxiv.org/abs/1705.03925} {arXiv:1705.03925 [astro-ph.CO]}
  \BibitemShut {NoStop}%
\bibitem [{\citenamefont {Chiang}\ and\ \citenamefont
  {Slosar}(2018)}]{Chiang:2018xpn}%
  \BibitemOpen
  \bibfield  {author} {\bibinfo {author} {\bibfnamefont {C.-T.}\ \bibnamefont
  {Chiang}}\ and\ \bibinfo {author} {\bibfnamefont {A.}~\bibnamefont
  {Slosar}},\ }\href@noop {} {\  (\bibinfo {year} {2018})},\ \Eprint
  {http://arxiv.org/abs/1811.03624} {arXiv:1811.03624 [astro-ph.CO]}
  \BibitemShut {NoStop}%
\bibitem [{\citenamefont {Hart}\ and\ \citenamefont
  {Chluba}(2022)}]{Hart:2021kad}%
  \BibitemOpen
  \bibfield  {author} {\bibinfo {author} {\bibfnamefont {L.}~\bibnamefont
  {Hart}}\ and\ \bibinfo {author} {\bibfnamefont {J.}~\bibnamefont {Chluba}},\
  }\href {\doibase 10.1093/mnras/stab2777} {\bibfield  {journal} {\bibinfo
  {journal} {Mon. Not. Roy. Astron. Soc.}\ }\textbf {\bibinfo {volume} {510}},\
  \bibinfo {pages} {2206} (\bibinfo {year} {2022})},\ \Eprint
  {http://arxiv.org/abs/2107.12465} {arXiv:2107.12465 [astro-ph.CO]}
  \BibitemShut {NoStop}%
\bibitem [{\citenamefont {Chluba}\ and\ \citenamefont
  {Hart}(2023)}]{Chluba:2023xqj}%
  \BibitemOpen
  \bibfield  {author} {\bibinfo {author} {\bibfnamefont {J.}~\bibnamefont
  {Chluba}}\ and\ \bibinfo {author} {\bibfnamefont {L.}~\bibnamefont {Hart}},\
  }\href@noop {} {\  (\bibinfo {year} {2023})},\ \Eprint
  {http://arxiv.org/abs/2309.12083} {arXiv:2309.12083 [astro-ph.CO]}
  \BibitemShut {NoStop}%
\bibitem [{\citenamefont {Lynch}\ \emph {et~al.}(2024)\citenamefont {Lynch},
  \citenamefont {Knox},\ and\ \citenamefont {Chluba}}]{Lynch:2024hzh}%
  \BibitemOpen
  \bibfield  {author} {\bibinfo {author} {\bibfnamefont {G.~P.}\ \bibnamefont
  {Lynch}}, \bibinfo {author} {\bibfnamefont {L.}~\bibnamefont {Knox}}, \ and\
  \bibinfo {author} {\bibfnamefont {J.}~\bibnamefont {Chluba}},\ }\href
  {\doibase 10.1103/PhysRevD.110.083538} {\bibfield  {journal} {\bibinfo
  {journal} {Phys. Rev. D}\ }\textbf {\bibinfo {volume} {110}},\ \bibinfo
  {pages} {083538} (\bibinfo {year} {2024})},\ \Eprint
  {http://arxiv.org/abs/2406.10202} {arXiv:2406.10202 [astro-ph.CO]}
  \BibitemShut {NoStop}%
\bibitem [{\citenamefont {Lee}\ \emph {et~al.}(2023)\citenamefont {Lee},
  \citenamefont {Ali-Ha\"\i{}moud}, \citenamefont {Sch\"oneberg},\ and\
  \citenamefont {Poulin}}]{Lee:2022gzh}%
  \BibitemOpen
  \bibfield  {author} {\bibinfo {author} {\bibfnamefont {N.}~\bibnamefont
  {Lee}}, \bibinfo {author} {\bibfnamefont {Y.}~\bibnamefont
  {Ali-Ha\"\i{}moud}}, \bibinfo {author} {\bibfnamefont {N.}~\bibnamefont
  {Sch\"oneberg}}, \ and\ \bibinfo {author} {\bibfnamefont {V.}~\bibnamefont
  {Poulin}},\ }\href {\doibase 10.1103/PhysRevLett.130.161003} {\bibfield
  {journal} {\bibinfo  {journal} {Phys. Rev. Lett.}\ }\textbf {\bibinfo
  {volume} {130}},\ \bibinfo {pages} {161003} (\bibinfo {year} {2023})},\
  \Eprint {http://arxiv.org/abs/2212.04494} {arXiv:2212.04494 [astro-ph.CO]}
  \BibitemShut {NoStop}%
\bibitem [{\citenamefont {Smith}\ \emph {et~al.}(2021)\citenamefont {Smith},
  \citenamefont {Poulin}, \citenamefont {Bernal}, \citenamefont {Boddy},
  \citenamefont {Kamionkowski},\ and\ \citenamefont {Murgia}}]{Smith:2020rxx}%
  \BibitemOpen
  \bibfield  {author} {\bibinfo {author} {\bibfnamefont {T.~L.}\ \bibnamefont
  {Smith}}, \bibinfo {author} {\bibfnamefont {V.}~\bibnamefont {Poulin}},
  \bibinfo {author} {\bibfnamefont {J.~L.}\ \bibnamefont {Bernal}}, \bibinfo
  {author} {\bibfnamefont {K.~K.}\ \bibnamefont {Boddy}}, \bibinfo {author}
  {\bibfnamefont {M.}~\bibnamefont {Kamionkowski}}, \ and\ \bibinfo {author}
  {\bibfnamefont {R.}~\bibnamefont {Murgia}},\ }\href {\doibase
  10.1103/PhysRevD.103.123542} {\bibfield  {journal} {\bibinfo  {journal}
  {Phys. Rev. D}\ }\textbf {\bibinfo {volume} {103}},\ \bibinfo {pages}
  {123542} (\bibinfo {year} {2021})},\ \Eprint
  {http://arxiv.org/abs/2009.10740} {arXiv:2009.10740 [astro-ph.CO]}
  \BibitemShut {NoStop}%
\bibitem [{\citenamefont {Lewis}\ and\ \citenamefont
  {Challinor}(2006)}]{Lewis:2006fu}%
  \BibitemOpen
  \bibfield  {author} {\bibinfo {author} {\bibfnamefont {A.}~\bibnamefont
  {Lewis}}\ and\ \bibinfo {author} {\bibfnamefont {A.}~\bibnamefont
  {Challinor}},\ }\href {\doibase 10.1016/j.physrep.2006.03.002} {\bibfield
  {journal} {\bibinfo  {journal} {Phys. Rept.}\ }\textbf {\bibinfo {volume}
  {429}},\ \bibinfo {pages} {1} (\bibinfo {year} {2006})},\ \Eprint
  {http://arxiv.org/abs/astro-ph/0601594} {arXiv:astro-ph/0601594} \BibitemShut
  {NoStop}%
\bibitem [{\citenamefont {Ade}\ \emph {et~al.}(2016)\citenamefont {Ade} \emph
  {et~al.}}]{Planck:2015mym}%
  \BibitemOpen
  \bibfield  {author} {\bibinfo {author} {\bibfnamefont {P.~A.~R.}\
  \bibnamefont {Ade}} \emph {et~al.} (\bibinfo {collaboration} {Planck}),\
  }\href {\doibase 10.1051/0004-6361/201525941} {\bibfield  {journal} {\bibinfo
   {journal} {Astron. Astrophys.}\ }\textbf {\bibinfo {volume} {594}},\
  \bibinfo {pages} {A15} (\bibinfo {year} {2016})},\ \Eprint
  {http://arxiv.org/abs/1502.01591} {arXiv:1502.01591 [astro-ph.CO]}
  \BibitemShut {NoStop}%
\bibitem [{\citenamefont {Smith}\ \emph
  {et~al.}(2022{\natexlab{b}})\citenamefont {Smith}, \citenamefont {Poulin},\
  and\ \citenamefont {Simon}}]{Smith:2022iax}%
  \BibitemOpen
  \bibfield  {author} {\bibinfo {author} {\bibfnamefont {T.~L.}\ \bibnamefont
  {Smith}}, \bibinfo {author} {\bibfnamefont {V.}~\bibnamefont {Poulin}}, \
  and\ \bibinfo {author} {\bibfnamefont {T.}~\bibnamefont {Simon}},\
  }\href@noop {} {\  (\bibinfo {year} {2022}{\natexlab{b}})},\ \Eprint
  {http://arxiv.org/abs/2208.12992} {arXiv:2208.12992 [astro-ph.CO]}
  \BibitemShut {NoStop}%
\bibitem [{\citenamefont {{Smith}}\ \emph {et~al.}(2012)\citenamefont
  {{Smith}}, \citenamefont {{Hanson}}, \citenamefont {{LoVerde}}, \citenamefont
  {{Hirata}},\ and\ \citenamefont {{Zahn}}}]{2012JCAP...06..014S}%
  \BibitemOpen
  \bibfield  {author} {\bibinfo {author} {\bibfnamefont {K.~M.}\ \bibnamefont
  {{Smith}}}, \bibinfo {author} {\bibfnamefont {D.}~\bibnamefont {{Hanson}}},
  \bibinfo {author} {\bibfnamefont {M.}~\bibnamefont {{LoVerde}}}, \bibinfo
  {author} {\bibfnamefont {C.~M.}\ \bibnamefont {{Hirata}}}, \ and\ \bibinfo
  {author} {\bibfnamefont {O.}~\bibnamefont {{Zahn}}},\ }\href {\doibase
  10.1088/1475-7516/2012/06/014} {\bibfield  {journal} {\bibinfo  {journal}
  {\jcap}\ }\textbf {\bibinfo {volume} {2012}},\ \bibinfo {eid} {014} (\bibinfo
  {year} {2012})},\ \Eprint {http://arxiv.org/abs/1010.0048} {arXiv:1010.0048
  [astro-ph.CO]} \BibitemShut {NoStop}%
\bibitem [{\citenamefont {Hotinli}\ \emph {et~al.}(2022)\citenamefont
  {Hotinli}, \citenamefont {Meyers}, \citenamefont {Trendafilova},
  \citenamefont {Green},\ and\ \citenamefont {van Engelen}}]{Hotinli:2021umk}%
  \BibitemOpen
  \bibfield  {author} {\bibinfo {author} {\bibfnamefont {S.~C.}\ \bibnamefont
  {Hotinli}}, \bibinfo {author} {\bibfnamefont {J.}~\bibnamefont {Meyers}},
  \bibinfo {author} {\bibfnamefont {C.}~\bibnamefont {Trendafilova}}, \bibinfo
  {author} {\bibfnamefont {D.}~\bibnamefont {Green}}, \ and\ \bibinfo {author}
  {\bibfnamefont {A.}~\bibnamefont {van Engelen}},\ }\href {\doibase
  10.1088/1475-7516/2022/04/020} {\bibfield  {journal} {\bibinfo  {journal}
  {JCAP}\ }\textbf {\bibinfo {volume} {04}},\ \bibinfo {pages} {020} (\bibinfo
  {year} {2022})},\ \Eprint {http://arxiv.org/abs/2111.15036} {arXiv:2111.15036
  [astro-ph.CO]} \BibitemShut {NoStop}%
\bibitem [{\citenamefont {Louis}\ \emph {et~al.}(2025)\citenamefont {Louis}
  \emph {et~al.}}]{Louis:2025tst}%
  \BibitemOpen
  \bibfield  {author} {\bibinfo {author} {\bibfnamefont {T.}~\bibnamefont
  {Louis}} \emph {et~al.},\ }\href@noop {} {\  (\bibinfo {year} {2025})},\
  \Eprint {http://arxiv.org/abs/2503.14452} {arXiv:2503.14452 [astro-ph.CO]}
  \BibitemShut {NoStop}%
\bibitem [{\citenamefont {Calabrese}\ \emph {et~al.}(2025)\citenamefont
  {Calabrese} \emph {et~al.}}]{Calabrese:2025mza}%
  \BibitemOpen
  \bibfield  {author} {\bibinfo {author} {\bibfnamefont {E.}~\bibnamefont
  {Calabrese}} \emph {et~al.},\ }\href@noop {} {\  (\bibinfo {year} {2025})},\
  \Eprint {http://arxiv.org/abs/2503.14454} {arXiv:2503.14454 [astro-ph.CO]}
  \BibitemShut {NoStop}%
\bibitem [{\citenamefont {{Ali-Ha{\"\i}moud}}\ and\ \citenamefont
  {{Bird}}(2013)}]{2013MNRAS.428.3375A}%
  \BibitemOpen
  \bibfield  {author} {\bibinfo {author} {\bibfnamefont {Y.}~\bibnamefont
  {{Ali-Ha{\"\i}moud}}}\ and\ \bibinfo {author} {\bibfnamefont
  {S.}~\bibnamefont {{Bird}}},\ }\href {\doibase 10.1093/mnras/sts286}
  {\bibfield  {journal} {\bibinfo  {journal} {\mnras}\ }\textbf {\bibinfo
  {volume} {428}},\ \bibinfo {pages} {3375} (\bibinfo {year} {2013})},\ \Eprint
  {http://arxiv.org/abs/1209.0461} {arXiv:1209.0461 [astro-ph.CO]} \BibitemShut
  {NoStop}%
\bibitem [{\citenamefont {Takahashi}\ \emph {et~al.}(2012)\citenamefont
  {Takahashi}, \citenamefont {Sato}, \citenamefont {Nishimichi}, \citenamefont
  {Taruya},\ and\ \citenamefont {Oguri}}]{Takahashi:2012em}%
  \BibitemOpen
  \bibfield  {author} {\bibinfo {author} {\bibfnamefont {R.}~\bibnamefont
  {Takahashi}}, \bibinfo {author} {\bibfnamefont {M.}~\bibnamefont {Sato}},
  \bibinfo {author} {\bibfnamefont {T.}~\bibnamefont {Nishimichi}}, \bibinfo
  {author} {\bibfnamefont {A.}~\bibnamefont {Taruya}}, \ and\ \bibinfo {author}
  {\bibfnamefont {M.}~\bibnamefont {Oguri}},\ }\href {\doibase
  10.1088/0004-637X/761/2/152} {\bibfield  {journal} {\bibinfo  {journal}
  {Astrophys. J.}\ }\textbf {\bibinfo {volume} {761}},\ \bibinfo {pages} {152}
  (\bibinfo {year} {2012})},\ \Eprint {http://arxiv.org/abs/1208.2701}
  {arXiv:1208.2701 [astro-ph.CO]} \BibitemShut {NoStop}%
\bibitem [{\citenamefont {Smith}\ \emph {et~al.}(2003)\citenamefont {Smith},
  \citenamefont {Peacock}, \citenamefont {Jenkins}, \citenamefont {White},
  \citenamefont {Frenk}, \citenamefont {Pearce}, \citenamefont {Thomas},
  \citenamefont {Efstathiou},\ and\ \citenamefont {Couchmann}}]{Smith:2002dz}%
  \BibitemOpen
  \bibfield  {author} {\bibinfo {author} {\bibfnamefont {R.~E.}\ \bibnamefont
  {Smith}}, \bibinfo {author} {\bibfnamefont {J.~A.}\ \bibnamefont {Peacock}},
  \bibinfo {author} {\bibfnamefont {A.}~\bibnamefont {Jenkins}}, \bibinfo
  {author} {\bibfnamefont {S.~D.~M.}\ \bibnamefont {White}}, \bibinfo {author}
  {\bibfnamefont {C.~S.}\ \bibnamefont {Frenk}}, \bibinfo {author}
  {\bibfnamefont {F.~R.}\ \bibnamefont {Pearce}}, \bibinfo {author}
  {\bibfnamefont {P.~A.}\ \bibnamefont {Thomas}}, \bibinfo {author}
  {\bibfnamefont {G.}~\bibnamefont {Efstathiou}}, \ and\ \bibinfo {author}
  {\bibfnamefont {H.~M.~P.}\ \bibnamefont {Couchmann}} (\bibinfo
  {collaboration} {VIRGO Consortium}),\ }\href {\doibase
  10.1046/j.1365-8711.2003.06503.x} {\bibfield  {journal} {\bibinfo  {journal}
  {Mon. Not. Roy. Astron. Soc.}\ }\textbf {\bibinfo {volume} {341}},\ \bibinfo
  {pages} {1311} (\bibinfo {year} {2003})},\ \Eprint
  {http://arxiv.org/abs/astro-ph/0207664} {arXiv:astro-ph/0207664} \BibitemShut
  {NoStop}%
\bibitem [{\citenamefont {{Mead}}\ \emph {et~al.}(2015)\citenamefont {{Mead}},
  \citenamefont {{Peacock}}, \citenamefont {{Heymans}}, \citenamefont
  {{Joudaki}},\ and\ \citenamefont {{Heavens}}}]{2015MNRAS.454.1958M}%
  \BibitemOpen
  \bibfield  {author} {\bibinfo {author} {\bibfnamefont {A.~J.}\ \bibnamefont
  {{Mead}}}, \bibinfo {author} {\bibfnamefont {J.~A.}\ \bibnamefont
  {{Peacock}}}, \bibinfo {author} {\bibfnamefont {C.}~\bibnamefont
  {{Heymans}}}, \bibinfo {author} {\bibfnamefont {S.}~\bibnamefont
  {{Joudaki}}}, \ and\ \bibinfo {author} {\bibfnamefont {A.~F.}\ \bibnamefont
  {{Heavens}}},\ }\href {\doibase 10.1093/mnras/stv2036} {\bibfield  {journal}
  {\bibinfo  {journal} {\mnras}\ }\textbf {\bibinfo {volume} {454}},\ \bibinfo
  {pages} {1958} (\bibinfo {year} {2015})},\ \Eprint
  {http://arxiv.org/abs/1505.07833} {arXiv:1505.07833 [astro-ph.CO]}
  \BibitemShut {NoStop}%
\bibitem [{\citenamefont {Mead}\ \emph {et~al.}(2021)\citenamefont {Mead},
  \citenamefont {Brieden}, \citenamefont {Tr\"oster},\ and\ \citenamefont
  {Heymans}}]{Mead:2020vgs}%
  \BibitemOpen
  \bibfield  {author} {\bibinfo {author} {\bibfnamefont {A.}~\bibnamefont
  {Mead}}, \bibinfo {author} {\bibfnamefont {S.}~\bibnamefont {Brieden}},
  \bibinfo {author} {\bibfnamefont {T.}~\bibnamefont {Tr\"oster}}, \ and\
  \bibinfo {author} {\bibfnamefont {C.}~\bibnamefont {Heymans}},\ }\href
  {\doibase 10.1093/mnras/stab082} {\bibfield  {journal} {\bibinfo  {journal}
  {Mon. Not. Roy. Astron. Soc.}\ }\textbf {\bibinfo {volume} {502}},\ \bibinfo
  {pages} {1401} (\bibinfo {year} {2021})},\ \Eprint
  {http://arxiv.org/abs/2009.01858} {arXiv:2009.01858 [astro-ph.CO]}
  \BibitemShut {NoStop}%
\bibitem [{\citenamefont {Bigwood}\ \emph {et~al.}(2024)\citenamefont {Bigwood}
  \emph {et~al.}}]{DES:2024iny}%
  \BibitemOpen
  \bibfield  {author} {\bibinfo {author} {\bibfnamefont {L.}~\bibnamefont
  {Bigwood}} \emph {et~al.} (\bibinfo {collaboration} {DES}),\ }\href {\doibase
  10.1093/mnras/stae2100} {\bibfield  {journal} {\bibinfo  {journal} {Mon. Not.
  Roy. Astron. Soc.}\ }\textbf {\bibinfo {volume} {534}},\ \bibinfo {pages}
  {655} (\bibinfo {year} {2024})},\ \Eprint {http://arxiv.org/abs/2404.06098}
  {arXiv:2404.06098 [astro-ph.CO]} \BibitemShut {NoStop}%
\bibitem [{\citenamefont {Gerbino}\ \emph {et~al.}(2020)\citenamefont
  {Gerbino}, \citenamefont {Lattanzi}, \citenamefont {Migliaccio},
  \citenamefont {Pagano}, \citenamefont {Salvati}, \citenamefont {Colombo},
  \citenamefont {Gruppuso}, \citenamefont {Natoli},\ and\ \citenamefont
  {Polenta}}]{Gerbino:2019okg}%
  \BibitemOpen
  \bibfield  {author} {\bibinfo {author} {\bibfnamefont {M.}~\bibnamefont
  {Gerbino}}, \bibinfo {author} {\bibfnamefont {M.}~\bibnamefont {Lattanzi}},
  \bibinfo {author} {\bibfnamefont {M.}~\bibnamefont {Migliaccio}}, \bibinfo
  {author} {\bibfnamefont {L.}~\bibnamefont {Pagano}}, \bibinfo {author}
  {\bibfnamefont {L.}~\bibnamefont {Salvati}}, \bibinfo {author} {\bibfnamefont
  {L.}~\bibnamefont {Colombo}}, \bibinfo {author} {\bibfnamefont
  {A.}~\bibnamefont {Gruppuso}}, \bibinfo {author} {\bibfnamefont
  {P.}~\bibnamefont {Natoli}}, \ and\ \bibinfo {author} {\bibfnamefont
  {G.}~\bibnamefont {Polenta}},\ }\href {\doibase 10.3389/fphy.2020.00015}
  {\bibfield  {journal} {\bibinfo  {journal} {Front. in Phys.}\ }\textbf
  {\bibinfo {volume} {8}},\ \bibinfo {pages} {15} (\bibinfo {year} {2020})},\
  \Eprint {http://arxiv.org/abs/1909.09375} {arXiv:1909.09375 [astro-ph.CO]}
  \BibitemShut {NoStop}%
\bibitem [{\citenamefont {Atkins}\ \emph {et~al.}(2023)\citenamefont {Atkins}
  \emph {et~al.}}]{Atkins:2023yzu}%
  \BibitemOpen
  \bibfield  {author} {\bibinfo {author} {\bibfnamefont {Z.}~\bibnamefont
  {Atkins}} \emph {et~al.},\ }\href {\doibase 10.1088/1475-7516/2023/11/073}
  {\bibfield  {journal} {\bibinfo  {journal} {JCAP}\ }\textbf {\bibinfo
  {volume} {11}},\ \bibinfo {pages} {073} (\bibinfo {year} {2023})},\ \Eprint
  {http://arxiv.org/abs/2303.04180} {arXiv:2303.04180 [astro-ph.CO]}
  \BibitemShut {NoStop}%
\bibitem [{\citenamefont {Brinckmann}\ \emph {et~al.}(2019)\citenamefont
  {Brinckmann}, \citenamefont {Hooper}, \citenamefont {Archidiacono},
  \citenamefont {Lesgourgues},\ and\ \citenamefont
  {Sprenger}}]{Brinckmann:2018owf}%
  \BibitemOpen
  \bibfield  {author} {\bibinfo {author} {\bibfnamefont {T.}~\bibnamefont
  {Brinckmann}}, \bibinfo {author} {\bibfnamefont {D.~C.}\ \bibnamefont
  {Hooper}}, \bibinfo {author} {\bibfnamefont {M.}~\bibnamefont
  {Archidiacono}}, \bibinfo {author} {\bibfnamefont {J.}~\bibnamefont
  {Lesgourgues}}, \ and\ \bibinfo {author} {\bibfnamefont {T.}~\bibnamefont
  {Sprenger}},\ }\href {\doibase 10.1088/1475-7516/2019/01/059} {\bibfield
  {journal} {\bibinfo  {journal} {JCAP}\ }\textbf {\bibinfo {volume} {01}},\
  \bibinfo {pages} {059} (\bibinfo {year} {2019})},\ \Eprint
  {http://arxiv.org/abs/1808.05955} {arXiv:1808.05955 [astro-ph.CO]}
  \BibitemShut {NoStop}%
\bibitem [{\citenamefont {Prabhu}\ \emph {et~al.}(2024)\citenamefont {Prabhu}
  \emph {et~al.}}]{SPT-3G:2024qkd}%
  \BibitemOpen
  \bibfield  {author} {\bibinfo {author} {\bibfnamefont {K.}~\bibnamefont
  {Prabhu}} \emph {et~al.} (\bibinfo {collaboration} {SPT-3G}),\ }\href
  {\doibase 10.3847/1538-4357/ad5ff1} {\bibfield  {journal} {\bibinfo
  {journal} {Astrophys. J.}\ }\textbf {\bibinfo {volume} {973}},\ \bibinfo
  {pages} {4} (\bibinfo {year} {2024})},\ \Eprint
  {http://arxiv.org/abs/2403.17925} {arXiv:2403.17925 [astro-ph.CO]}
  \BibitemShut {NoStop}%
\bibitem [{\citenamefont {Adame}\ \emph {et~al.}(2025)\citenamefont {Adame}
  \emph {et~al.}}]{DESI:2024mwx}%
  \BibitemOpen
  \bibfield  {author} {\bibinfo {author} {\bibfnamefont {A.~G.}\ \bibnamefont
  {Adame}} \emph {et~al.} (\bibinfo {collaboration} {DESI}),\ }\href {\doibase
  10.1088/1475-7516/2025/02/021} {\bibfield  {journal} {\bibinfo  {journal}
  {JCAP}\ }\textbf {\bibinfo {volume} {02}},\ \bibinfo {pages} {021} (\bibinfo
  {year} {2025})},\ \Eprint {http://arxiv.org/abs/2404.03002} {arXiv:2404.03002
  [astro-ph.CO]} \BibitemShut {NoStop}%
\end{thebibliography}%

\appendix
\section{Impact of non-linear structure formation}\label{app:nonlinear}
It is well known that non-linear structure formation enhances the matter power spectrum on the smallest scales. However, such an increased clustering also necessarily changes the gravitational potential, which in turns causes stronger lensing of the smallest scales CMB anisotropies. Therefore, in order to perform an accurate CMB analysis of scales with $\ell \gtrsim 3000$ (where the spectrum becomes dominated by the lensing effects, see also \cite{Lewis:2006fu}) it is crucial to be aware of and model this impact of non-linear structure formation on the lensed CMB anistropies. In our case we use the simple \texttt{halofit} prediction from \cite{2013MNRAS.428.3375A,Takahashi:2012em} based on the initial work of \cite{Smith:2002dz}. However, there are also other prescriptions to estimate the impact of non-linear clustering, such as from the HMcode \cite{2015MNRAS.454.1958M,Mead:2020vgs} framework. We take the difference between the lensed power spectra for the bestfit $\Lambda$CDM model using these two approaches as an estimate of the order of magnitude of the impact of the non-linear modeling uncertainties. More importantly, the HMcode framework also allows us to estimate the impact of baryonic feedback using their baryonification model. In our case we simply evaluate the differences between the predictions with strong and without baryonic feedback as an estimate of the uncertainty due to baryonic feedback. We choose $T_\mathrm{AGN}=10^8\mathrm{Kelvin}$, which can be considered a conservative estimate of the associated uncertainties. However, we do note that some recent observations also point towards strong baryonic feedback \cite{DES:2024iny}.

\section{Predicted uncertainties}\label{app:uncertainties}
We estimate the uncertainties of near-future CMB data using a simple approach based on the simple Gaussian approximation from \cite{Gerbino:2019okg} with beam noise
\begin{equation}
    N^{i,XX}_\ell = (\epsilon^{XX}_\ell)^2 \exp[- \ell (\ell + 1) \sigma^{i,XX}_\mathrm{FWHM}/(8 \ln 2)]
\end{equation}
where we take the values of $\sigma^{i,XX}_\mathrm{FWHM}$ for each channel from \cite[Tab.~1]{Atkins:2023yzu} for ACT and \cite[Tab.~1]{Brinckmann:2018owf} for CMB-S4, adding the different channels using the inverse-variance weighted mean. We use $f_\mathrm{sky}\approx 0.4$ for both, and $\epsilon^{TT}_\ell = 15\mu \mathrm{K \cdot arcmin}$, $\epsilon^{EE}_\ell = 20\mu \mathrm{K \cdot arcmin}$ for ACT and \cite[Tab.~1]{Brinckmann:2018owf} for CMB-S4. Explicitly the covariance is modeled using \cite[Eq.~(5)]{SPT-3G:2024qkd}.

\section{Other likelihoods}\label{app:likelihoods}
\begin{figure*}
    \centering
    \includegraphics[width=0.45\linewidth]{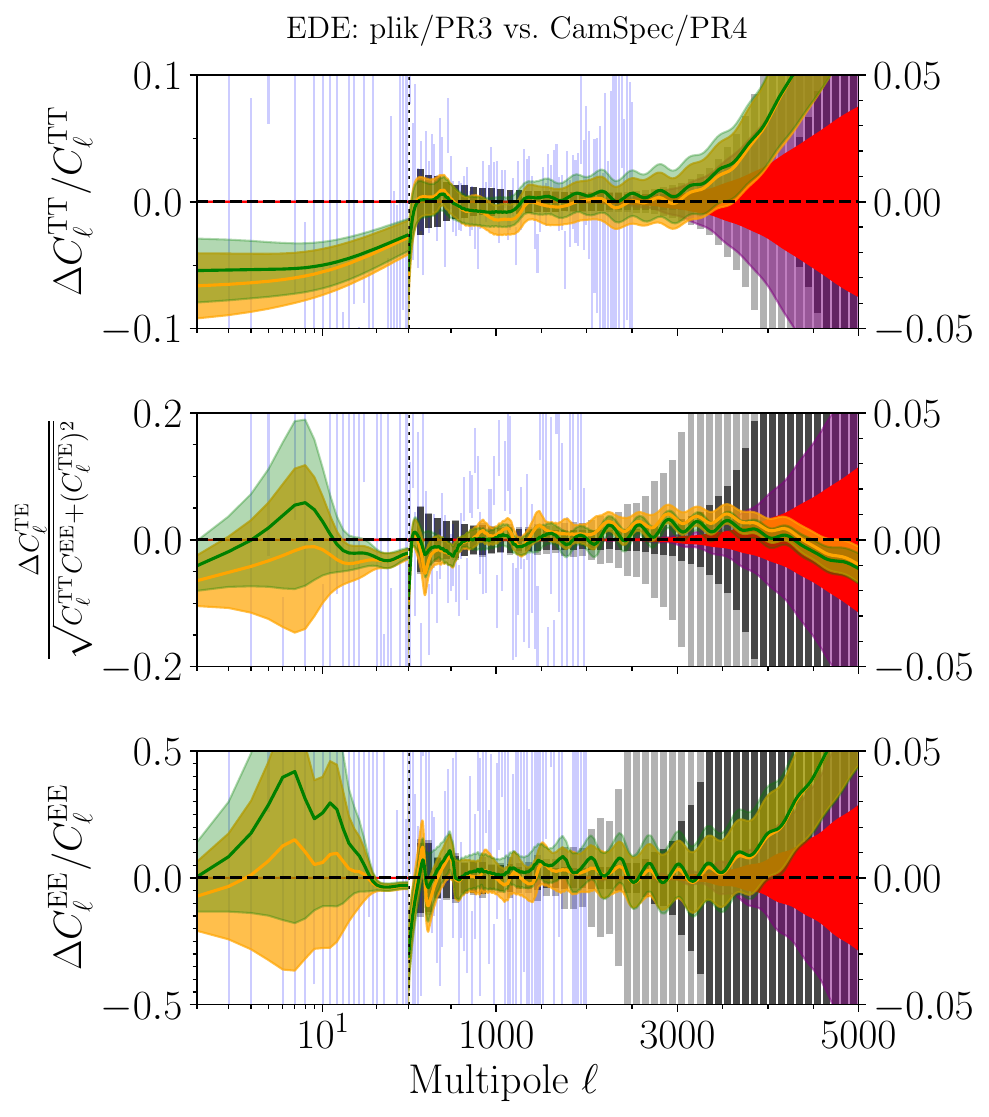}
    \includegraphics[width=0.45\linewidth]{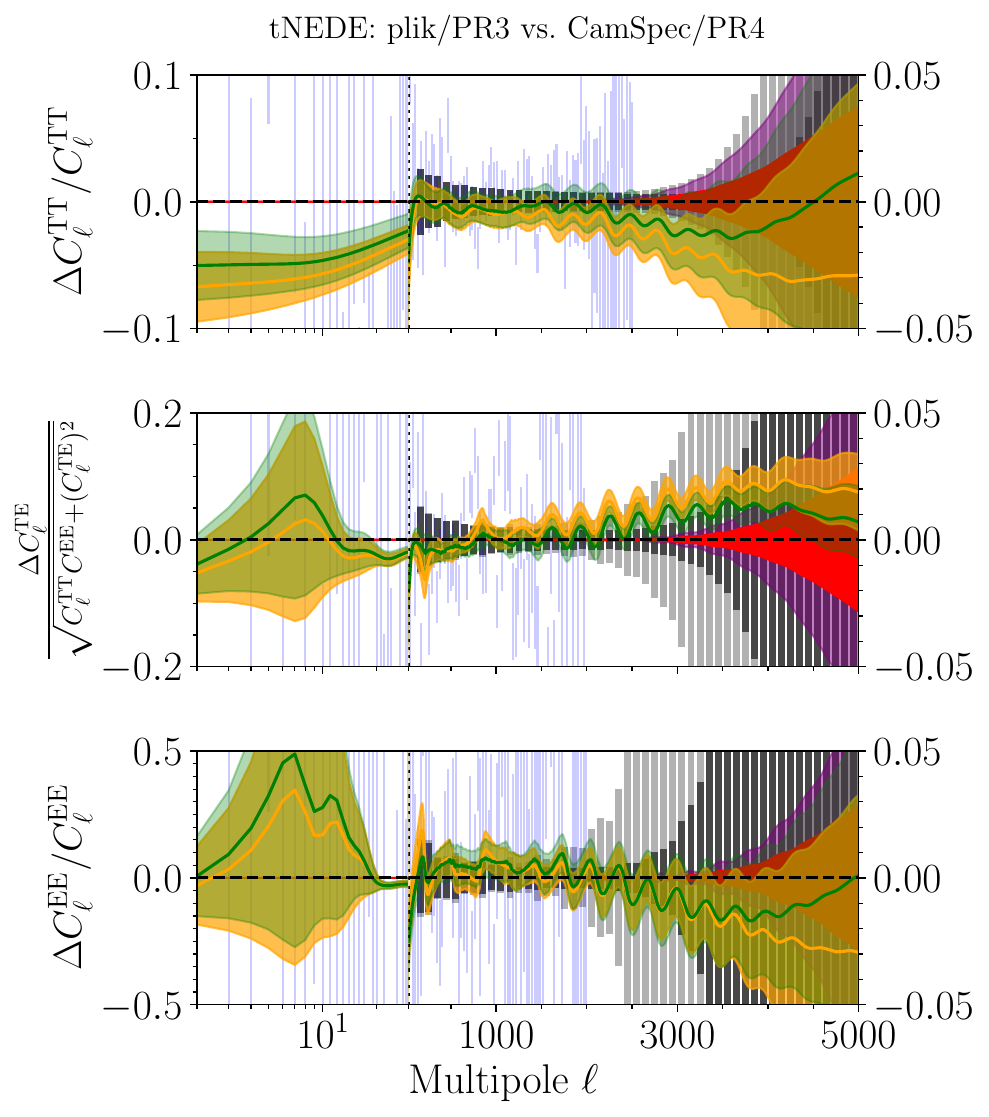} \\
    \includegraphics[width=0.45\linewidth]{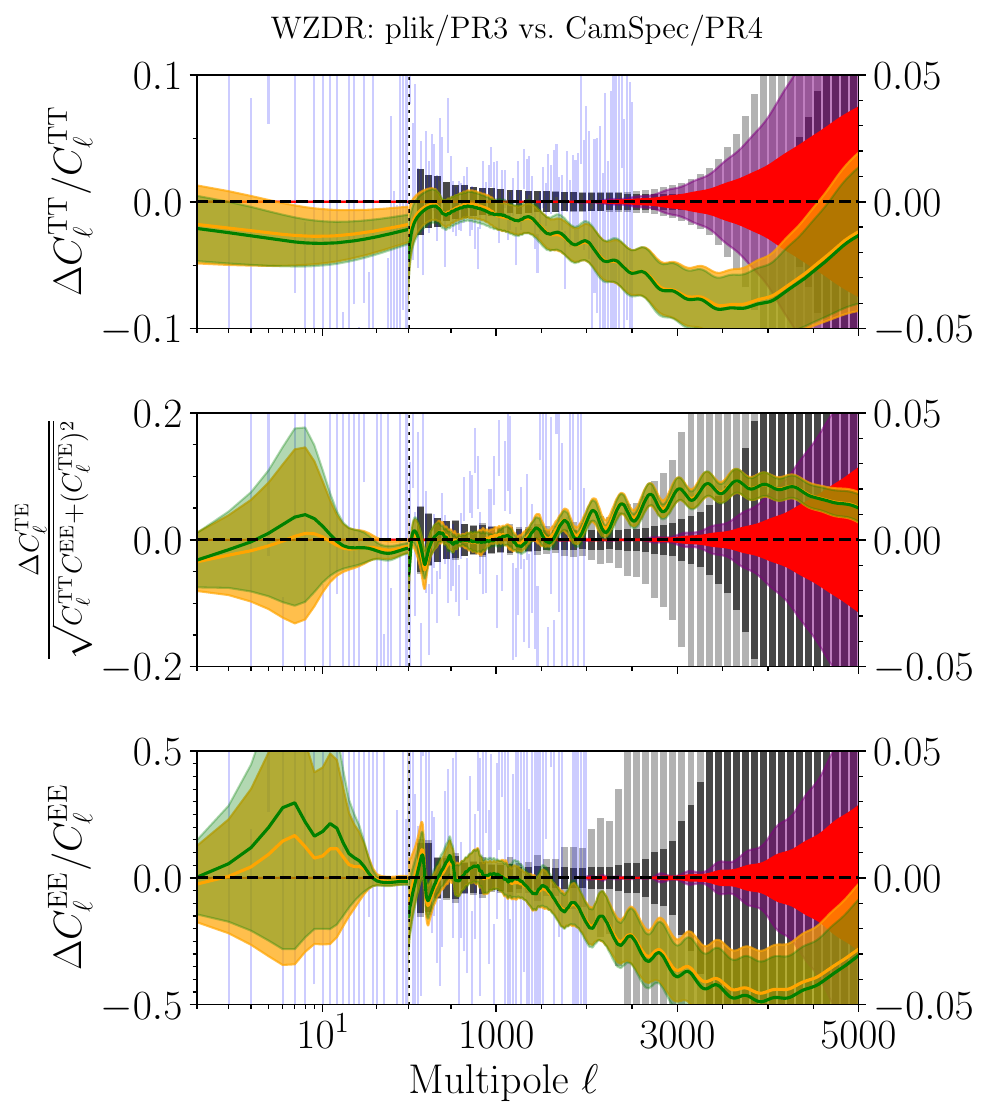}
    \includegraphics[width=0.45\linewidth]{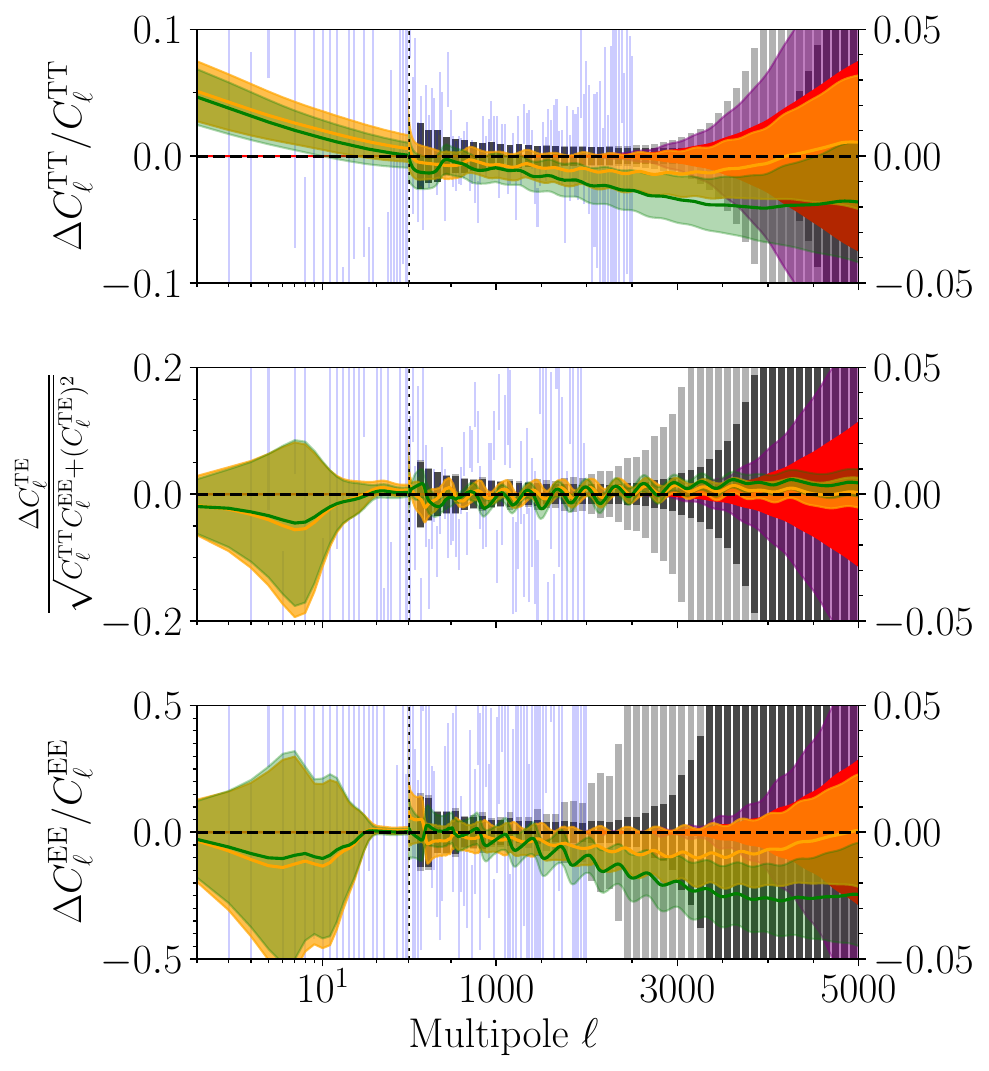}
    \caption{As in \cref{fig:wzdr} but comparing the \texttt{plik} predictions (orange) to the \texttt{CamSpec} predictions (green). They are all compared to the same baseline \LCDM reference from \texttt{plik}. Top left: EDE. Top right: tNEDE. Bottom left: WZDR. Bottom right: Different value of the electron mass in the early universe.}
    \label{fig:likelihoods}
\end{figure*}

We can also wonder if the measurability of a given model depends on with which CMB likelihood we compute the initial MCMC chain from which the predictions are drawn. We compare the \texttt{plik} and \texttt{CamSpec} results for the EDE, tNEDE, WZDR, and the varying $m_e$ models in \cref{fig:likelihoods} for the same reference model obtained with the \texttt{plik} likelihood. The results are largely the same and overall very consistent.

As a final check, for the WZDR model we explicitly confirmed that using the BAO likelihoods advertised in the main text and using the DESI-Y1 BAO likelihoods from Ref.~\cite{DESI:2024mwx} doesn't make a significant difference.

\end{document}